\documentclass[11pt, a4paper,onecolumn]{belarticle4}
\pdfoutput=1
\usepackage{revsymb, amsmath, amsfonts, amssymb, enumerate, fullpage, amsthm, graphicx, braket, relsize, bbm, mathrsfs, mathtools}
\usepackage{graphicx,color}
\definecolor{darkbrown}{rgb}{0.4, 0.26, 0.13}
\definecolor{ao}{rgb}{0.0, 0.5, 0.0}
\definecolor{bleudefrance}{rgb}{0.19, 0.55, 0.91}
\usepackage{adjustbox}

\usepackage{paralist}
\usepackage{subfigure}

\usepackage[linktocpage=true, colorlinks=true, linkcolor=blue, urlcolor=blue, citecolor=blue]{hyperref}

\usepackage{cite}

\usepackage{tcolorbox}
\tcbuselibrary{skins,breakable}
\usetikzlibrary{shadings,shadows}

    {\endtcolorbox}

\begin{document}

\title{Theory-agnostic nonclassicality certification in an integrated photonic circuit}

\author{Vinicius P.~Rossi$^*$}
\email{vinicius.pretti-rossi@ug.edu.pl}
\affiliation{International Centre for Theory of Quantum Technologies, University of  Gda{\'n}sk, 80-309 Gda{\'n}sk, Poland} 

\author{Emanuele Polino$^*$}
\affiliation{Queensland Quantum and Advanced Technologies Research Institute, Griffith  University, Yuggera Country,   Brisbane,  Queensland,  4111  Australia}

\author{Beatrice Polacchi}
\affiliation{Dipartimento di Fisica - Sapienza Universit\`{a} di Roma, P.le Aldo Moro 5, I-00185 Roma, Italy}
\affiliation{\textit{Current affiliation: }Sparrow Quantum, Nordre Fasanvej 215, DK-2000 Frederiksberg, Denmark}

\author{Valeria Cimini} 
\affiliation{Dipartimento di Fisica - Sapienza Universit\`{a} di Roma, P.le Aldo Moro 5, I-00185 Roma, Italy}

\author{David Schmid}
\affiliation{Perimeter Institute for Theoretical Physics, Waterloo, Ontario, Canada, N2L 2Y5}

\author{John H.~Selby}
\affiliation{International Centre for Theory of Quantum Technologies, University of  Gda{\'n}sk, 80-309 Gda{\'n}sk, Poland}

\author{Giacomo Corrielli}
\affiliation{Istituto di Fotonica e Nanotecnologie, Consiglio Nazionale delle Ricerche (IFN-CNR), Piazza Leonardo da Vinci, 32, I-20133 Milano, Italy}

\author{Andrea Crespi}
\affiliation{Dipartimento di Fisica, Politecnico di Milano, Piazza Leonardo da Vinci, 32, I-20133 Milano, Italy}
\affiliation{Istituto di Fotonica e Nanotecnologie, Consiglio Nazionale delle Ricerche (IFN-CNR), Piazza Leonardo da Vinci, 32, I-20133 Milano, Italy}

\author{Roberto Osellame}
\affiliation{Istituto di Fotonica e Nanotecnologie, Consiglio Nazionale delle Ricerche (IFN-CNR), Piazza Leonardo da Vinci, 32, I-20133 Milano, Italy}

\author{Fabio Sciarrino} 
\affiliation{Dipartimento di Fisica - Sapienza Universit\`{a} di Roma, P.le Aldo Moro 5, I-00185 Roma, Italy}

\author{Ana Belén Sainz}
\affiliation{International Centre for Theory of Quantum Technologies, University of  Gda{\'n}sk, 80-309 Gda{\'n}sk, Poland} 
\affiliation{Basic Research Community for Physics e.V., Germany}

\begingroup
\renewcommand{\thefootnote}{*}
\footnotetext{These authors share first authorship.}
\endgroup

\begin{abstract}
Certifying nonclassicality without assuming any particular underlying physical theory constitutes both a foundational challenge and a requirement for device-independent protocols. Realizing such theory-independent certification in scalable architectures is essential for connecting fundamental tests with emerging quantum technologies. Integrated photonic circuits are a leading platform for scalable quantum information processing, motivating the development of rigorous methods to certify the nonclassical resources underpinning quantum advantage.
In this paper, we report a theory-independent certification of generalized contextuality on a three-mode integrated photonic circuit. Our approach combines theory-agnostic tomography with simplex-embeddability certification—a framework that requires no assumptions about the underlying physical theory—and applies it to experimental data from a three-mode photonic circuit seeded by single photons.
An independently constructed quantum model of the experiment provides a consistency check on the observed nonclassicality, without entering the certification as a theoretical assumption. The method rules out simplex-embeddability for our experimental data, providing robust, theory-independent evidence that our photonic circuit exhibits nonclassicality.
\end{abstract}

\maketitle

\newpage

\tableofcontents

\section{Introduction}\label{sec:intro}

The power of quantum technologies ultimately rests on physical behaviors that have no classical counterpart. Such nonclassicality underpins a broad range of quantum tasks, including computation~\cite{nielsen2000}, communication~\cite{teleport}, cryptography~\cite{cryptography}, randomness certification~\cite{random}, and metrology~\cite{circuit2}, among others~\cite{horodecki,acin18}. Establishing and certifying these nonclassical resources is therefore central not only to the foundations of quantum theory, but also to the development of scalable quantum technologies.
Generalized contextuality arguably emerges as a prime notion of nonclassicality~\cite{spekkens2005contextuality}. At its core, generalized noncontextuality requires operationally equivalent experimental procedures, i.e., procedures that are indistinguishable by all operational statistics, to admit identical representations in any underlying ontological model~\cite{spekkens2005contextuality}. Built on the foundational works of J.S. Bell~\cite{bell64} and S. Kochen and E. P. Specker~\cite{ks}, generalized contextuality can simultaneously be connected to a plethora of other notions of nonclassicality~\cite{rossi23,SEER,schmid18ineq,wright2023invertible,kunjwal2019anomalous,naim24,NEGATIVITY,Schmid2024structuretheorem, walleghem2025,sina25} and pinpointed as the source of quantum advantage in multiple tasks~\cite{spekkens09,saha2019,sumit23, roch22,fonseca25,schmid22,bowles23, ambainis19, chailloux16, yadavalli22,lostaglio20,lostaglio22,schmid2018contextual,shin21,flatt22,mukherjee22, yile25}, while being generally available in quantum setups~\cite{tipi}. It is therefore a philosophically, foundationally, and practically well-motivated criterion of nonclassicality. Recent work established the equivalence between the existence of a noncontextual ontological model for an operational theory and the possibility of simplex embedding for the associated generalized probabilistic theory~\cite{schmid2021simplex}. A simplex embedding amounts to representing the states and effects of the GPT within a classical simplex, whose vertices correspond to underlying classical states, and its existence can be tested via linear programming~\cite{selby2024linear, cavalcanti2023github}.

There is strong motivation, however, for developing assessments of nonclassicality that are valid independently of the quantum formalism. On the one hand, device-independent approaches, which do not rely on the inner workings of the employed apparatuses~\cite{scarani2012device,poderini2022ab}, enable the development of protocols that remain cryptographically secure even when the devices are untrusted. On the other hand, this paradigm ensures that the certified advantage survives even if quantum theory is eventually superseded or deemed inadequate. For the case of contextuality, device-independent-like assessments can be pursued through a technique called theory-agnostic tomography. 
Here, one can use the experimental data itself to construct the building blocks of a physical theory (not necessarily quantum), which allows one to analyze the statistical data and assess its classical character while making minimal assumptions~\cite{mazurek2021experimentally,grabowecky2022experimentally}. 
This approach naturally requires notions of nonclassicality that are not themselves intrinsically quantum, a requirement that generalized contextuality conveniently complies with by being defined purely operationally~\cite{spekkens2005contextuality}. Together, theory-agnostic tomography and simplex-embeddability certification constitute a methodological protocol for asserting nonclassicality without assuming quantum theory at any stage---gather data from an experiment, construct a physical theory with minimal assumptions that reproduces the data, and verify whether this theory is nonclassical.

This programme has so far remained experimentally limited. Theory-agnostic simplex-embeddability assessments have been demonstrated on a superconducting qubit~\cite{aloy2024theory}, and variations of this approach appear on a photonic qubit~\cite{mazurek2021experimentally} (with inequality-based, rather than simplex-embeddability, assessments of contextuality) and a photonic qutrit~\cite{grabowecky2022experimentally}, without implementing any assessment of contextuality after the theory-agnostic reconstruction. Hitherto, no demonstration of a complete theory-independent certification method has been reported beyond qubit-scale systems or on integrated photonic platforms. This gap is significant because photonics, and in particular integrated photonic circuits ~\cite{circuit1,wang2020integrated}, are among the leading architectures for scalable quantum technologies. Integrated photonics combines phase stability, miniaturization, precision, reconfigurability, and compatibility with established fabrication methods, making it a natural platform in which to connect foundational certification protocols with realistic quantum devices and tasks such as quantum metrology~\cite{circuit2, caruccio2025quantum, valeri2020experimental} and quantum computation~\cite{zhu2026recent}.

In this work, we implement theory-agnostic tomography and simplex-embeddability certification in a three-mode integrated photonic circuit realized via the femtosecond laser writing technique and capable of performing quantum information processing~\cite{corrielli2021femtosecond}. The experiment implements a prepare-and-measure scenario in which a single photon is injected into the circuit and detected at one of three output modes, while independently controlled optical phases define the preparation and measurement procedures. By varying the phases used to realize the preparation and measurement stages of the experiment, we obtain a data table of conditional output probabilities. Rather than assuming a Hilbert-space model for these data, we first use theory-agnostic tomography to infer the effective operational state and effect spaces directly from the measured statistics. A rank analysis identifies the dimension of the reconstructed generalized probabilistic fragment that best captures the data without overfitting. We then apply simplex-embeddability certification to the reconstructed fragment, thereby testing whether the observed statistics admits any classical noncontextual explanation. The experimentally reconstructed fragment rules out any simplex-embeddability with nonzero robustness, certifying nonclassicality without assuming quantum theory. An independently constructed quantum model, used only as a consistency check, reproduces the same effective rank and yields a comparable certification, whereas a classical counterpart is correctly identified as classical.

Our results establish theory-independent certification of nonclassicality in an integrated photonic platform and extend the full theory-agnostic tomography and simplex-embeddability pipeline beyond qubit-scale systems. They show that integrated photonics can support not only quantum information-processing tasks, but also operational certification protocols that make minimal reference to an underlying physical theory.

\section{Preliminaries}\label{sec:prelim}

We now present the operational framework by which nonclassical behaviour can be certified without assuming the validity of quantum theory. First, let us remark that we aim to make these claims based on \textit{observational data}: we prepare a physical system according to a chosen procedure, perform a measurement, and record an outcome (usually called a prepare-and-measure scenario); we care about the outcomes that we observe, and how these depend on the choice of measurements and preparation procedures. In particular, we do not assume any physical theory as the explanation of our experimental observations --- the data are obtained without any commitment to an underlying theory.

We then want to draw conclusions on whether our observations resist a classical explanation, based only on the statistical behaviour of our experimental data. In the following subsections, we provide the basics (and further references) on how nonclassicality can be certified from this statistical behaviour --- importantly, in a theory-agnostic way.

\begin{figure}[t]
\centering
\includegraphics[width=0.7\linewidth]{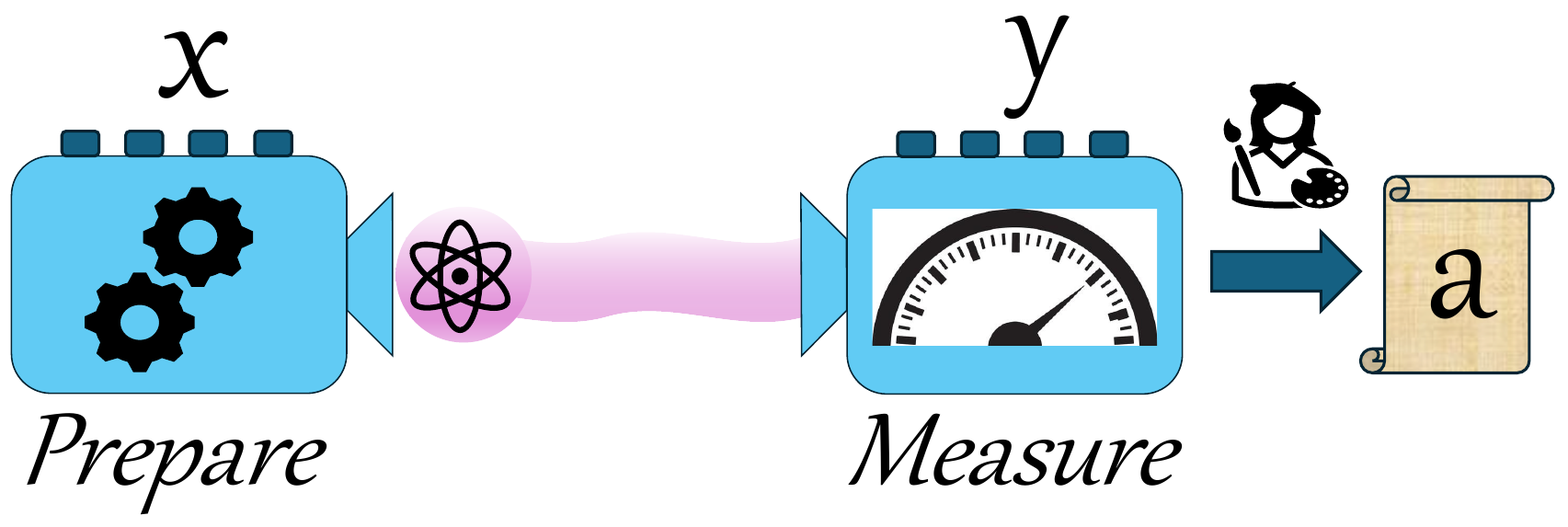}
    \caption{Conceptual scheme of a prepare-and-measure scenario.}
    \label{fig:PM}
\end{figure}

\subsection{Nonclassicality certification}\label{sec:NC}

The scenario we are interested in is the so-called \emph{prepare-and-measure scenario} briefly mentioned above. Let us denote by $x$ the classical label of the \textit{preparation procedures} that we might use to prepare our system (\textit{states}), by $y$ the classical label of the \textit{measurement procedures} that we may implement on our system, and by $a$ the classical label of the measurement outcome that we obtain. In this work, $x$, $y$, and $a$ are discrete variables that may take values from finite sets. A schematic representation of a generic prepare-and-measure scenario is provided in Fig.~\ref{fig:PM}. 
The \textit{conditional probability distribution} (also referred to as \textit{correlations} or \textit{behaviour}) that captures the statistical data in the experiment is given by
\begin{align}\label{eq:objPM}
\{p(a|y,x)\}_{a,y,x}\,.
\end{align}

The question then is: given a behaviour in the form of Eq.~\eqref{eq:objPM}, how can we decide whether or not it admits a classical explanation? Here, the notion of classicality that we endorse is that of generalized noncontextuality (hereafter simply \textit{noncontextuality}) \cite{spekkens2005contextuality}. At its core, generalized noncontextuality is the requirement that experimental procedures that are operationally indistinguishable (i.e., procedures that yield identical statistics in all operational contexts) be represented identically at the ontological level.

A recent geometrical underpinning of noncontextuality \cite{schmid2021simplex} frames the question of the existence of a noncontextual ontological model within the framework of generalized probabilistic theories (GPTs) \cite{hardy2001quantum,barrett2007information}. Here, one seeks to find a set of states $\mathcal{S} = \{s_x\}$ (i.e., a set of vectors in a real vector space) and a set of effects $\mathcal{E} = \{e_{a|y}\}$ (i.e., a set of co-vectors in the dual real vector space), that satisfy certain properties. On the one hand, we want  $\mathcal{S}$ and $\mathcal{E}$ to be tomographically complete for each other\footnote{In the traditional jargon of generalized contextuality \cite{spekkens2005contextuality}, this constraint is what ensures that the so-called operational equivalences are satisfied.}. On the other hand, we want these states and effects to reproduce the statistical data: 
\begin{align}\label{eq:PMgpt}
    p(a|y,x) = e_{a|y}[s_x]\,.
\end{align}

If one can find a pair $\mathcal{S},\mathcal{E}$ with those properties, and which is moreover \textit{simplex-embeddable}, then the data table is deemed classically explainable \cite{schmid2021simplex,Schmid2024structuretheorem}. 

Given a pair $\mathcal{S},\mathcal{E}$, checking simplex-embeddability is an instance of a linear program \cite{selby2024linear}, and there are software repositories tailored to this task \cite{selby2024linear,cavalcanti2023github}: these codes determine whether the pair is simplex-embeddable and, if it is not, return the weight of a partially depolarising channel acting on the states that suffices to make the pair simplex-embeddable (see Appendix~\ref{app:codes}).

\subsection{Theory-agnostic tomography}\label{sec:TA-tomo}

Theory-agnostic tomography for preparations and measurements was first introduced in Ref.~\cite{mazurek2021experimentally}, where it was applied to a qubit in a photonic experiment. The idea is that we have a data table given by the conditional probability distribution $\{p(a|y,x)\}_{a,y,x}$ (hereafter referred to as the \textit{data table}) in a prepare-and-measure scenario, and we do not want to assume that there is a quantum description of our experiment. We would still want to represent our experiment as having preparation and measurement procedures, but with a minimal underpinning so we can reason about their statistical properties. That is, we want to represent what happens in our experiment using the formalism of GPTs. The problem then is: given a data table $D$ with entries $D_{x,a|y}:=p(a|y,x)$, how can we find a real vector space, and subsets of vectors therein, so that corresponding sets of states $\mathcal{S}$ and effects $\mathcal{E}$ can be identified which reproduce the observed statistics via Eq.~\eqref{eq:PMgpt}? One key challenge is identifying the dimension that the real vector space should have.

Theory-agnostic tomography selects the best-fitting dimension as the one which least overfits the data (for example, by means of the Akaike information criterion (AIC) score \cite{akaike1974new}); it then finds the sets of states $\mathcal{S}$ and effects $\mathcal{E}$ therein which reproduce the desired statistics. This technique has been applied in many experiments (see, e.g., Refs.~\cite{grabowecky2022experimentally,aloy2024theory}), and there are software repositories tailored to performing theory-agnostic tomography \cite{aloy2024github}.

The standard theory-agnostic tomography analysis consists of, given a data table $D$ and a rank $k$, finding a factorization $S^T\cdot E$ that best describes $D$, such that the columns of $S$ and $E$ are valid states/effects for a GPT in $\mathbb{R}^k$. The decision of which model best fits $D$ is made by optimizing the parameter
\begin{equation}\label{eq:chi}
    \chi^2(k,D) :=\min_{S,E}\sum_{ij}\frac{(S_i^T\cdot E_j-D_{ij})^2}{(\Delta D_{ij})^2}\,,
\end{equation}
where $S_i$ and $E_j$ are the $i$-th column of $S$ and $j$-th column of $E$, respectively, and $(\Delta D_{ij})^2$ is the variance associated with the frequency provided by entry $D_{ij}$. This variance is estimated by assuming that $D$ was obtained from a finite number of iterations $N$ via multinomial sampling (see Ref.~\cite{aloy2024theory} for a detailed explanation). Estimating $\chi^2(k,D)$ is a bilinear program, and local minima can be found via a seesaw algorithm~\cite{aloy2024github}.

This procedure is sufficient only when the rank $k$ is known. Otherwise, one must investigate which rank best describes the experimental data. In this work, we search for $k$ by analyzing a collection of data tables---all corresponding to the same experiment---which may differ slightly due to statistical fluctuations. We thus have two different statistical quantities, namely $\chi^2_{train}$, the average of the optimal $\chi^2$ in Eq.~\ref{eq:chi} obtained for independently generated data tables; and $\chi_{test}^2$, obtained by comparing how well the optimal $S$ and $E$ for a particular data table describes the other data tables, and search for the rank $k$ that minimizes both parameters simultaneously. In Appendix~\ref{ap:rank} we present the details of the method we use to find the best $k$. 

\section{The Experimental Setup}\label{sec:setup}

\subsection{Description of the interferometric setup in the integrated circuit}

Our experimental setup consists of an integrated photonic chip~\cite{polino2019experimental,valeri2020experimental,cimini2021calibration} with three input ports and three output ports, implementing the prepare-and-measure scenario as depicted in Fig.~\ref{fig:tritter}-A. In each round of the experiment, we input a photon through the first mode of the interferometer and measure through which mode it exits.
In this setup, the preparation and measurements correspond to setting the relative phases of the circuit. This physical encoding has been used in the prepare-and-measure scenario to test the nonclassicality of the delayed-choice experiment~\cite{chaves2018causal,polino2019device,yu2019realization}.
More specifically, there are four relative phases that can be tuned during the experiment: a first pair ($\phi^x_{1},\phi^x_{2}$) represents the parameters of the preparation stage, and a second pair ($\phi^y_{1},\phi^y_{2}$) represents those of the measurement stage. We study the output statistics for different values of these interferometer phases, with the interferometer seeded by single photons. The two pairs of phases are changed by tuning the values of two voltages: $\Delta V_{x}$ and $\Delta V_{y}$, respectively. 

Experimental data are obtained by independently varying the voltages $\Delta V_x$ between $0$ and $55.1$~mV and $\Delta V_y$ between $0$ and $52.1$~mV, with a photon always entering the circuit through the first mode of the interferometer. For each output, data are collected as coincidence counts generated by a spontaneous parametric down-conversion (SPDC) source, in which one photon is used as a herald to reduce background noise. We register, on average, around 3500 counts per voltage configuration at the chip's output ports; these counts constitute the data. Figs.~\ref{fig:tritter}-B--D show the collected experimental statistics as black dots.

We want to certify the nonclassicality of the collected statistics by understanding the experiment as a prepare-and-measure one, where we think of $\Delta V_{x}$ (resp.~$\Delta V_{y}$) as tuning only  $\phi^x_{1}$ and $\phi^x_{2}$ (resp.~$\phi^y_{1}$ and $\phi^y_{2}$) which hence can be thought of as defining the preparation stage (resp.~measurement stage). In the remainder of this section (together with Appendix \ref{app:models} and the detailed depiction of Fig.~\ref{fig:circuit-detailed}), we discuss how this can indeed be the case.

\subsection{Theoretical model for the integrated photonic chip}

\begin{figure}[t]
    \centering    
    A)\adjustbox{valign=t}{\includegraphics[width=0.6\linewidth]{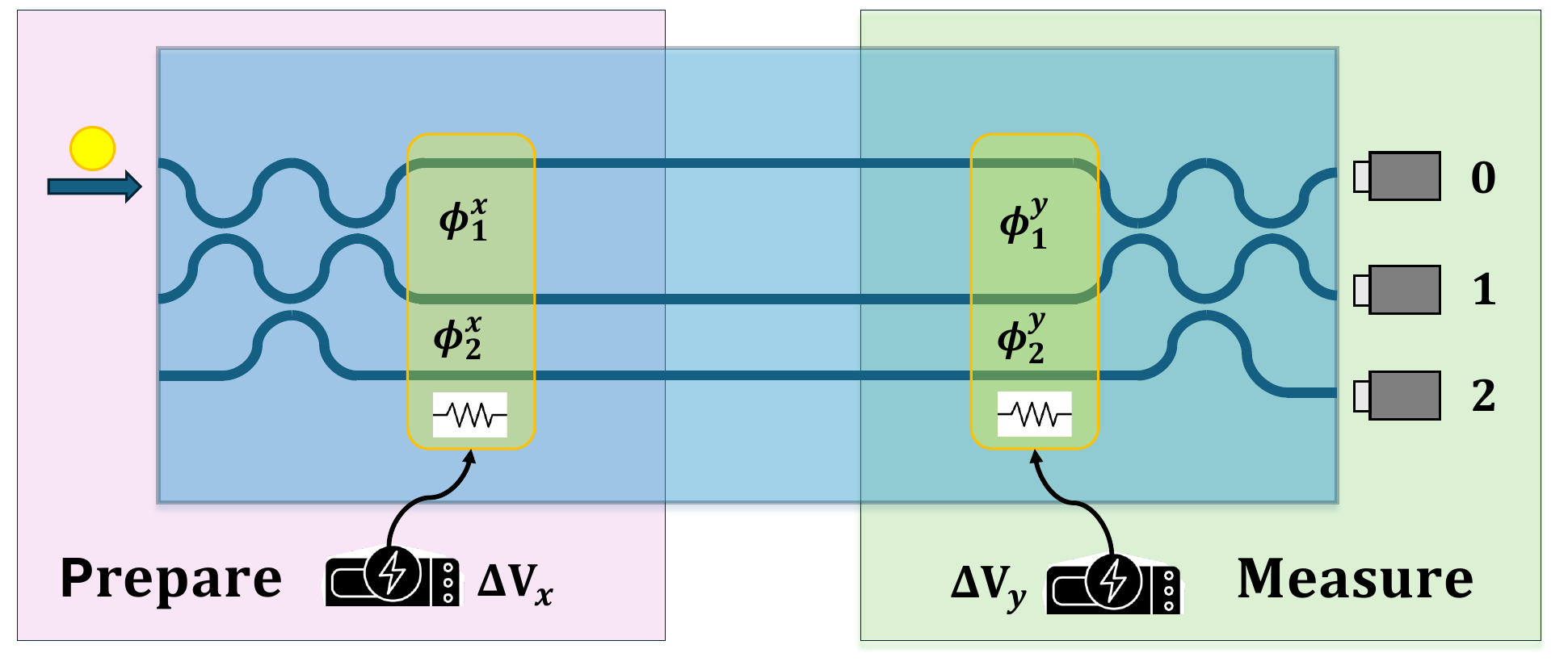}}\\
    \begin{tabular}{ccc}
    B)\adjustbox{valign=t}{\includegraphics[width=0.29\linewidth]{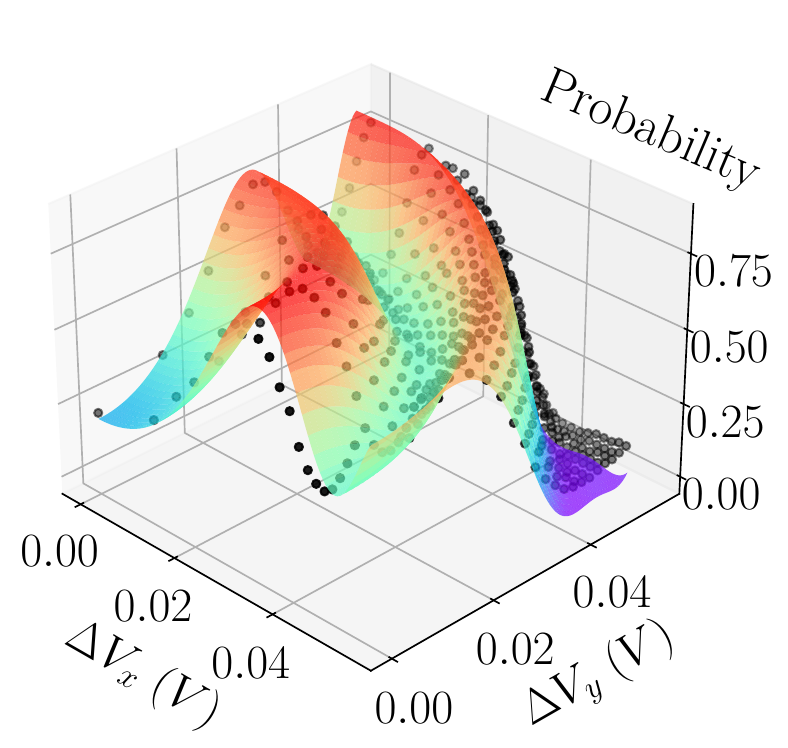}} & C)\adjustbox{valign=t}{\includegraphics[width=0.29\linewidth]{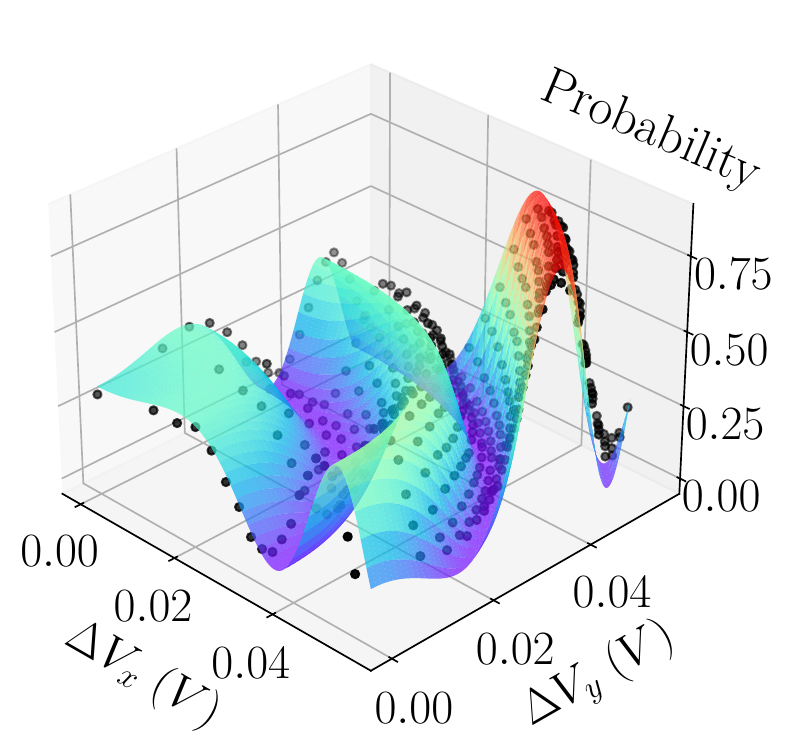}} & D)\adjustbox{valign=t}{\includegraphics[width=0.29\linewidth]{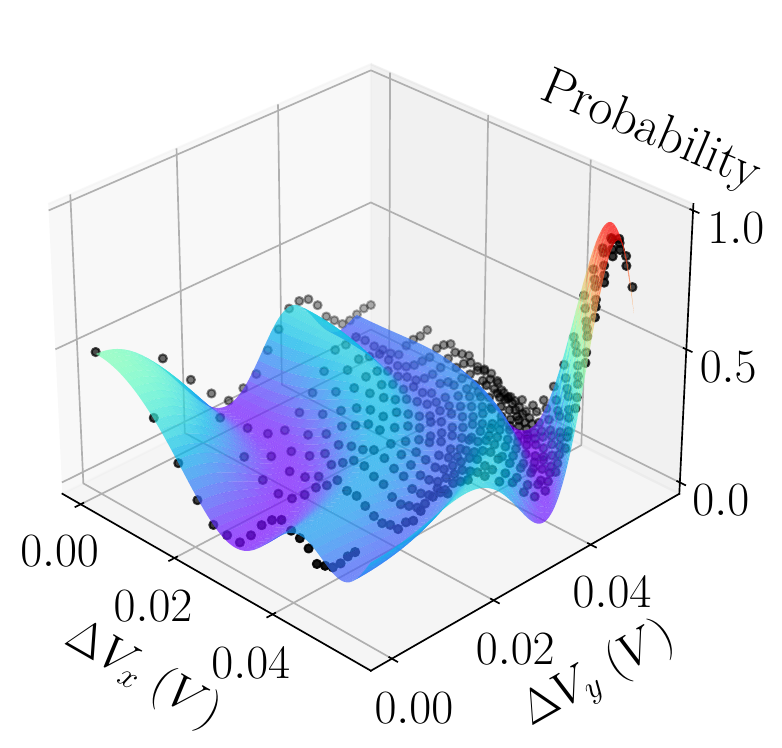}}
    \end{tabular}
    \caption{A) Scheme of the experimental setup involving a single photon. B--D) Probabilities for the theoretical model of a photon entering the setup through the first mode of the interferometer and being detected in (B) mode $0$, (C) mode $1$, and (D) mode $2$, as a function of the experimental voltages $\Delta V_x$ and $\Delta V_y$. Points scattered in the plots represent experimental frequencies, while the 3D function is the theoretical quantum model of the circuit's output probabilities.}
    \label{fig:tritter}
\end{figure}

 For validation purposes, we construct a quantum model that approximates the experimental statistics (see Appendix~\ref{app:models}). The model captures preparation and measurement procedures through response functions for the preparation phases $\phi^x_{1(2)}$ and the measurement phases $\phi^y_{1(2)}$. These are functions of the dissipated powers $P(\Delta V_{x})$ and $P(\Delta V_{y})$ generated respectively by the preparation voltage $\Delta V_{x}$ and the measurement voltage $\Delta V_{y}$, and take the form

\begin{alignat}{2}
    \phi^x_{1(2)}(\Delta V_x)&=\alpha_{x\,1(2)} P(\Delta V_x)+\alpha_{x\,1(2)2}P^2(\Delta V_x)\,,\\
    \phi^y_{1(2)}(\Delta V_y)&=\alpha_{y\,1(2)}P(\Delta V_y)+\alpha_{y\,1(2)2}P^2(\Delta V_y)\,.
\end{alignat}
The total physical phases that the photons experience along the interferometer are then $\phi_{1}=\phi^x_{1}(\Delta V_x)+\phi^y_{1}(\Delta V_y)+\phi_{01}$ and $\phi_{2}=\phi^x_{2}(\Delta V_x)+\phi^y_{2}(\Delta V_y)+\phi_{02}$, where $\phi_{01(2)}$ are constant phase shifts characterizing the circuit when no voltage is applied to the resistors.
Figures~\ref{fig:tritter}-B--D display how the theoretical model (coloured surface) accommodates the experimental data (black dots). The corresponding theoretical data table is constructed by passing the same values of $\Delta V_x$ and $\Delta V_y$ to the theoretical model, and therefore has the same shape as the experimental data table.

Our analysis builds on a prepare-and-measure description of the experiment, in which the experimental procedure naturally decomposes into a preparation stage followed by a measurement stage. This structure is supported by several independent features of the implementation. First, the parameters $\Delta V_x$ and $\Delta V_y$ are independently controlled and act, respectively, as preparation and measurement settings. Second, the theoretical model reproducing the experimental statistics admits the same decomposition at the level of the phase shifts. Finally, as discussed below, the theory-agnostic analysis yields the same optimal factorization rank for both the theoretical probabilities and the experimental data. Together, these observations provide strong evidence that the prepare-and-measure scenario accurately captures the causal structure of the experiment, thereby justifying the factorization of the data matrix into a set of states and a set of effects. Further details are provided in Appendix~\ref{app:models}.

\subsection{A note on loopholes}

Just like with experimental certifications of other forms of nonclassicality, assessments of generalized contextuality over GPT fragments reconstructed via theory-agnostic tomography are also subject to loopholes. In particular, it has been established that these fragments constitute so-called \emph{GPT shadows}, which may break assessments of nonclassicality based on simplex-embeddability (such as the one employed in this work)~\cite{gitton24,schmid2025shadows}. While assessments of classicality are trustworthy, a generic feature of any experimental contextuality certification is that there is no way to fully ensure that the nonclassicality proofs obtained are loophole-free, unless some assumptions are made or justified\footnote{For instance, if one assumes that quantum theory underpins the experiment, and moreover trusts the preparation and measurement devices, then GPT shadows do not open a loophole, as they become equivalent to the GPT fragment itself.}. 
What becomes relevant in our specific experiment is the local quantum dimension of the system under study. We can argue that a natural conceptualisation for the system under investigation is a three-level quantum system, which, along with the theory-agnostic analysis for the optimal dimension of the (theoretical and experimental) shadows, establishes an upper bound on the dimension of the true GPT from which the shadow is constructed.\footnote{This does not rule out arguments conjecturing that, despite the experimentally and statistically justified assumptions, the experimenter might still be inadvertently ignoring certain degrees of freedom of an ultimately classical system. Since we cannot ensure relative tomography in this experiment~\cite{schmid2025shadows}, we can never rule out conspiratorial arguments of this type.}  Once this upper bound is justified, results such as the one in Ref.~\cite{lidia19} may assist in ruling out whether the derived shadow is problematic for the nonclassicality assessment. Although those specific results do not translate directly to our current problem, we investigate in a coming work how to estimate the number of degrees of freedom one has to miss to run into untrustworthy assessments.

\section{Results}\label{sec:results}

\subsection{Rank analysis}\label{se:rank}

Here we present the results on the GPT dimension that best suits our experimental data. For this, we perform rank analysis not only of the given data, but also of fluctuations of this experimental data table generated by multinomial sampling. We moreover perform rank analysis on the ideal data table that arises from our analytical model of the integrated photonic circuit, together with additional (theoretical) data tables derived from the ideal one via multinomial sampling, assuming a finite total number of counts. We find the optimal rank for our experiment by studying the ranks of all these data tables.

We start by performing the rank analysis on the data table obtained from the theoretical model. To do so, we generate the central data table numerically and additionally sample 9 data tables that are deviations from the central data table via multinomial sampling over the outcomes with the theoretical frequencies, and a total number of coincidences assumed to be $N=10^6$.  The choice of working with 10 data tables in total provides 90 comparisons for the estimation of $\chi_{\mathrm{test}}^2$, which is sufficient to obtain smooth estimates of the error while keeping computational costs moderate. The selected value for $N$ is an artificial number, chosen to be much larger than the average number of coincidences obtained experimentally (of the order of $N\approx 3.5\cdot10^3$), and is relevant both for the sampling of ancillary data tables and for the seesaw algorithm in Ref.~\cite{aloy2024github}, which is devised for experimental data and requires frequencies obtained from a finite number of counts. Moreover, we further show in Appendix~\ref{app:sanity} that the rank analysis for these theoretical data tables is stable for multinomial sampling with a total number of coincidences $N\geq10^5$. The numerical analysis of $\chi_{\mathrm{train}}^2$ provided in Ref.~\cite{aloy2024github} consists of a seesaw algorithm that will converge within a given tolerance value, which we set as $\epsilon=10^{-3}$ for the analysis of the theoretical model, as this is the minimal tolerance value that yielded convergence for all generated tables. We also perform this analysis for the experimental frequencies, generating 9 additional data tables via multinomial sampling using the experimental number of coincidences for each setting $(\Delta V_x,\Delta V_y)$. We then implement the algorithm from Ref.~\cite{aloy2024github} with a tolerance $\epsilon=10^{-6}$.

\begin{figure}
    \centering
    \begin{tabular}{cc}
    A)\adjustbox{valign=t}{\includegraphics[width=0.45\linewidth]{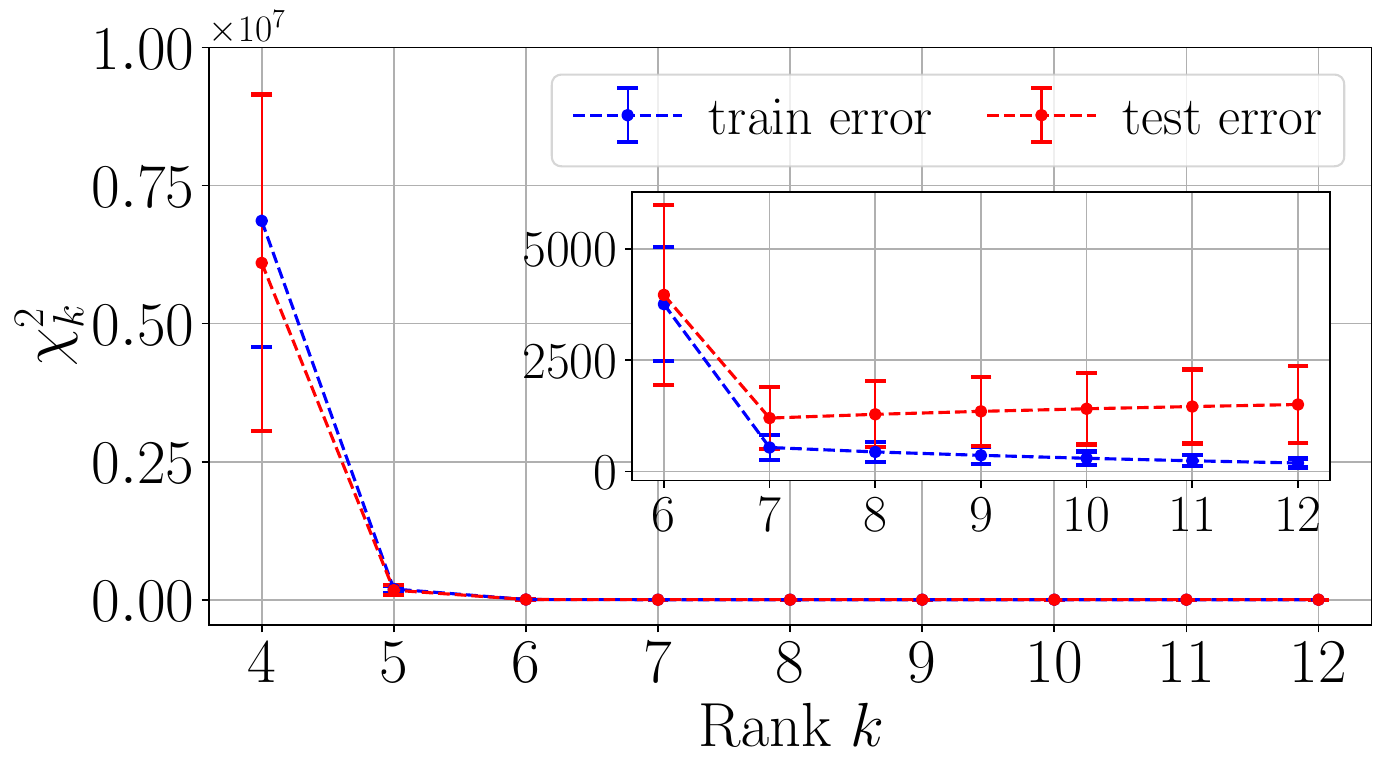}}
    & 
    B)\adjustbox{valign=t}{\includegraphics[width=0.45\linewidth]{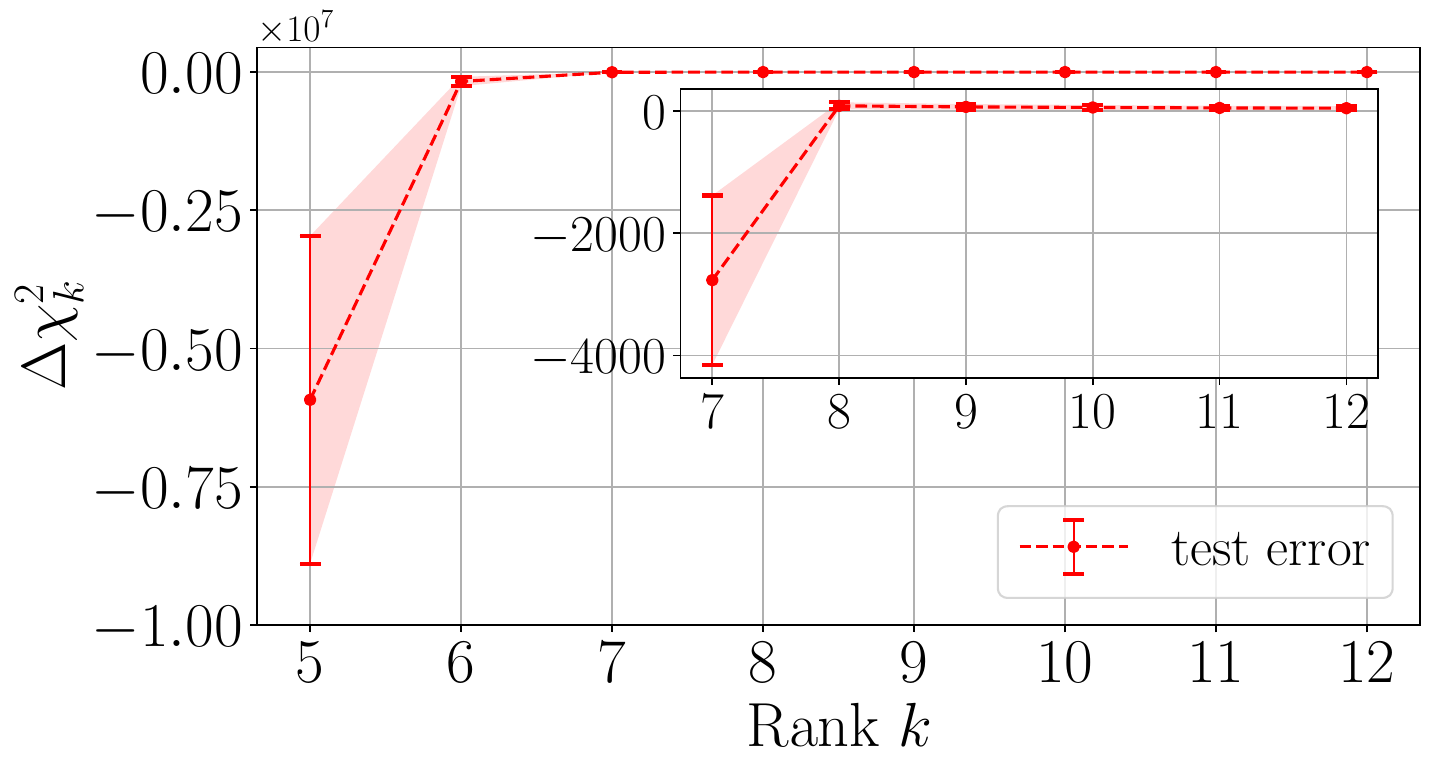}}\\
    C)\adjustbox{valign=t}{\includegraphics[width=0.45\linewidth]{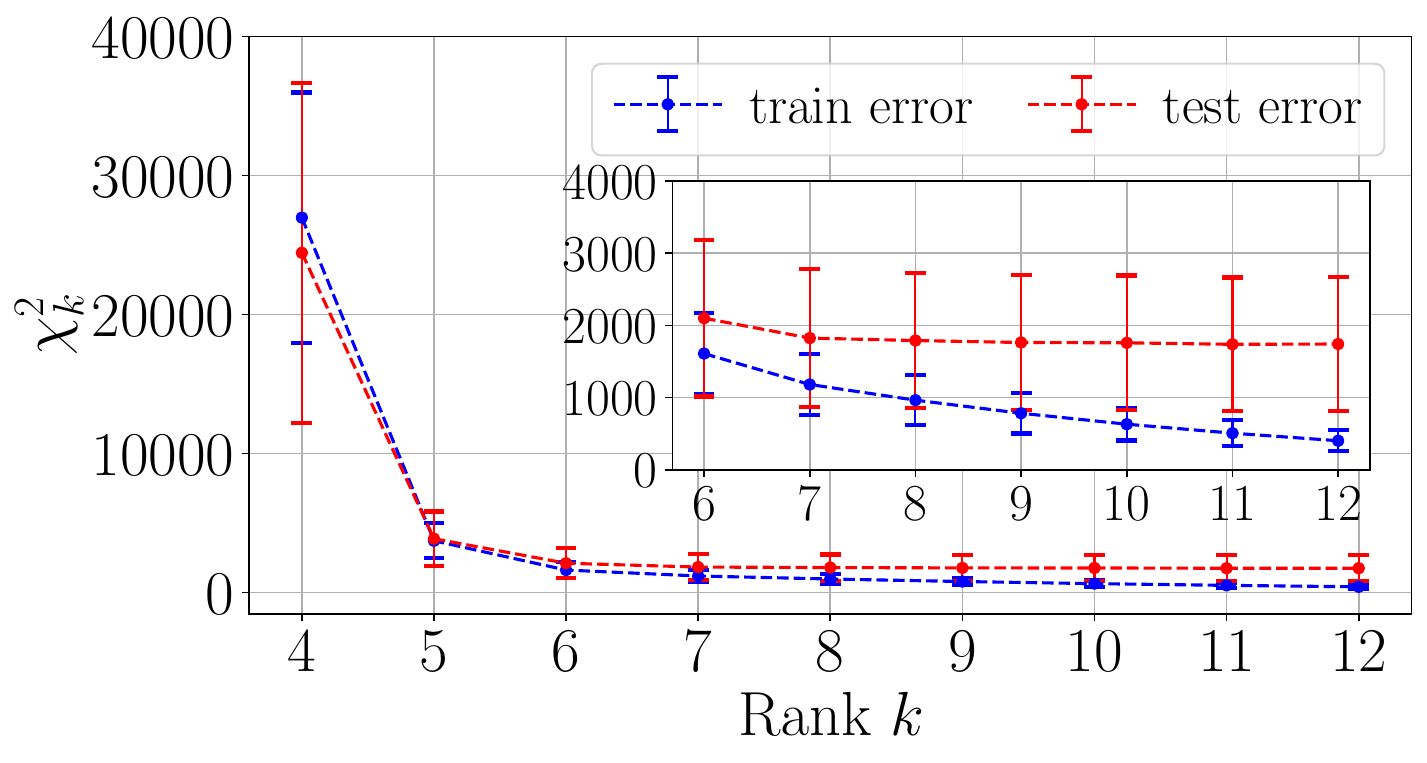}} & D)\adjustbox{valign=t}{\includegraphics[width=0.45\linewidth]{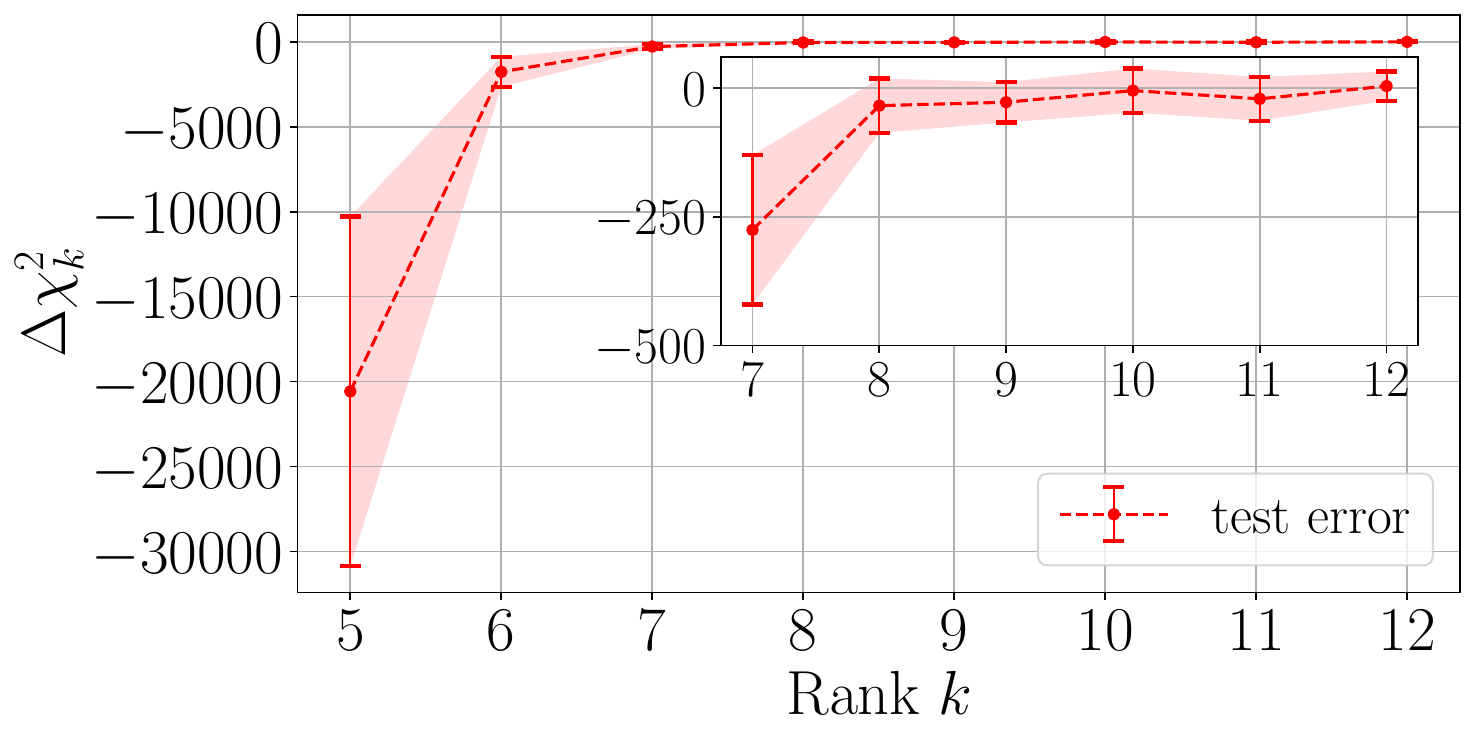}}
    \end{tabular}
    \caption{Rank analysis for (A--B) the theoretical model and (C--D) the experimental data. (A,C) $\chi^2_{\mathrm{train}}$ and $\chi^2_{\mathrm{test}}$ as functions of the rank $k$ of the data table factorization; (B,D) change in $\chi^2_{\mathrm{test}}$ as the rank increases from $k-1$ to $k$. The optimal rank is the highest value of $k$ such that this change in $\chi^2_{\mathrm{test}}$ is still negative.}
    \label{fig:chitheo2}
\end{figure}

Figure~\ref{fig:chitheo2} summarises the results for (A--B) the theoretical model and (C--D) the experimental data. Figs.~\ref{fig:chitheo2}-A and \ref{fig:chitheo2}-C show how $\chi^2_{\mathrm{train}}$ and $\chi^2_{\mathrm{test}}$ change with the rank $k$ assumed for the factorization of the data table.  One can see that the assumption of $k=4$ yields a much larger value for both $\chi^2_{\mathrm{train}}$ and $\chi^2_{\mathrm{test}}$, which quickly collapses for $k=5$. This holds for both the theoretical probabilities and the experimental data. Figs.~\ref{fig:chitheo2}-B and \ref{fig:chitheo2}-D portray the change in $\chi^2_{\mathrm{test}}$ as the rank increases, and the optimal rank for the factorization will be the highest value of $k$ for which this change is still negative. For both theoretical and experimental data tables, we see that when $k$ jumps from 7 to 8, the change in $\chi^2_{\mathrm{test}}$ becomes positive (within one standard deviation, for the experimental case). This shows that the rank analyses for the theoretical model and the experimental data yield the same optimal rank for the factorization, $k_{\mathrm{optimal}}=7$.

Given that the theoretical model for the setup is a quantum 3-level system, one would naturally expect $k_{\mathrm{optimal}}=9$ (at least for the theoretical data), since the GPT dimension of a quantum system of dimension $d$ is $d^2$. The reason why theory-agnostic tomography finds a smaller rank can be traced back to the structure of the experimental setup itself, as either the preparation or the measurement stage does not span the full GPT space. Indeed, once we consider all the states generated by the theoretical model, we see that the dimension of their linear span is equal to 7, and not 9. This is also why this model is not tomographically complete: although the set of all projectors spans the full space of Hermitian qutrit operators and can therefore properly capture all equivalences between the states, the states cannot perfectly distinguish some inequivalent effects.

\subsection{Theory-agnostic tomography and nonclassicality assessment}

We can therefore perform theory-agnostic tomography on the data, following the seesaw algorithm from Ref.~\cite{aloy2024github}, assuming the factorization rank to be $k=7$. The output of the algorithm is the set of states and effects of a generalized probabilistic theory that realizes the experimental data. We then input these states and effects into a linear program developed to assess the contextuality of data in prepare-and-measure scenarios underpinned by any GPT~\cite{selby2024linear,cavalcanti2023github}. 

The key figure of merit studied by the program is the robustness $r$ quantifying how much depolarising noise the reconstructed GPT states can tolerate while remaining nonclassical. Specifically, it is the largest $r\in[0,1]$ such that replacing each state $S_i$ with its partially depolarised counterpart $(1-r)\cdot S_i+r\cdot\mu$, where $\mu$ is the maximally mixed state, still yields a nonclassical GPT fragment\footnote{In this work, the depolarising channel is constructed from the theory-agnostic unit effect and the centroid of the theory-agnostic states.}. It is important to note, however, that this should not be interpreted as meaning that the experiment would still demonstrate contextuality if one were to replace the data table $D$ by a depolarised version $(1-r')\cdot D+r'\cdot D_{\frac13}$, for any $r'<r$, where $D_{\frac13}$ is the data table with all entries equal to $\frac13$ (the uniform distribution over the three outcomes). This is because applying noise directly to the data table may lead to a different GPT description from theory-agnostic tomography, and potentially a different optimal rank. The robustness should instead be understood as a property of the particular GPT fragment reconstructed from the provided data: it certifies that this reconstructed fragment is itself robustly nonclassical, independently of how noise on the raw data might affect the tomographic reconstruction.

For the theory-agnostic output of the experimental data, the linear program finds the average robustness of contextuality to depolarising noise (over the original data table and all the sampled ones) to be $r = 0.200\pm0.005$.  This is the same value with or without normalizing counterparts for the effects\footnote{I.e., with or without including $u-e$ for every $e$ in the set of theory-agnostic effects, with $u$ the unit effect}, which means that the theory-agnostic output is likely capturing the normalization of effects. We do the same analysis for the theory-agnostic output of the data tables built from the theoretical model, obtaining an average robustness of $r = 0.261\pm0.001$. Once again, this value is similar to the one we obtain if we include the normalizing counterpart to the effects. The observation of nonzero robustness for both the experimental and theoretical reconstructions, together with the previously established bounds on the factorization rank and the rank analysis from theory-agnostic tomography, certifies the presence of nonclassicality in the integrated photonic circuit.

We note that this proof of contextuality, as well as any fully theory-independent demonstration, is subject to the tomographic completeness loophole. One could devise further experiments that minimize the likelihood that the loophole is exploited\footnote{E.g., an experiment where, by assuming a specific quantum description of the setup, the data are compatible with relatively-tomographically-complete states and measurements, as done in Ref.~\cite{aloy2024theory}.}, but from a fundamental viewpoint, this loophole will always exist with non-zero likelihood.

\section{Conclusions}\label{sec:conclusions}

We have demonstrated theory-agnostic certification of generalized contextuality in a three-mode integrated photonic circuit seeded by single photons. Starting from the observed prepare-and-measure statistics, we determined the effective dimension of the reconstructed operational theory without assuming any quantum description of the device. The statistical analysis identified an optimal rank of $k=7$ for both the experimental data and the independently constructed quantum model. Theory-agnostic tomography at this rank produced GPT fragments whose simplex-embeddability was then assessed through linear programming. For the experimental reconstruction, we obtained an average robustness to depolarisation of $0.200\pm0.005$, compared with $0.261\pm0.001$ for the theoretical reconstruction. By contrast, the corresponding analysis of the data generated by the fully classical counterpart described in Appendix \ref{app:sanity} returned a robustness of zero. Taken together, these results demonstrate that the generated data are incompatible with any generalized noncontextual ontological model.

To our knowledge, this work provides the first experimental demonstration on a photonic platform of the combined framework of theory-agnostic tomography and simplex-embeddability certification, and the first implementation of this approach beyond a qubit-scale system.\\
Previous experiments either addressed only part of this framework or relied on an explicitly quantum description of the devices~\cite{mazurek2021experimentally,grabowecky2022experimentally,aloy2024theory,rafa23}. Here, an effective GPT fragment is reconstructed directly from the observed statistics and tested for simplex-embeddability, without assuming a quantum representation of the preparations and measurements.

This framework occupies a unique position between model-dependent certification and fully device-independent Bell tests. It does not require a trusted quantum description of the preparations and measurements (which can be problematic in applications where security is required to hold even in the presence of imperfectly characterized or untrusted devices~\cite{toni07}), yet remains more experimentally accessible than the technologically demanding loophole-free Bell protocols~\cite{friedman18,metger21}. Moreover, by targeting generalized contextuality, it applies to experimental scenarios and potential resources beyond those captured by Bell nonclassicality~\cite{tipi}. While theory-agnostic nonclassicality can come with assumptions concerning tomographic completeness, it provides a practical route to certification when neither a detailed quantum model nor a fully device-independent test is suitable.

Our results also identify two directions for further investigation. First, it will be important to determine whether the certified nonclassicality can be converted into an operational advantage in a communication, computation, sensing, or interrogation task (for this, see Ref.~\cite{rafa23}). Theory-agnostic certification could then establish that the advantage is genuinely nonclassical without requiring the quantum formalism as a premise. Moreover, the fact that the optimal rank for theory-agnostic tomography may be smaller than the GPT dimension suggested by the underlying quantum system could help circumvent the conjectured exponential complexity of simplex-embeddability certification~\cite{sahandeh26complexity}, since the theory-agnostic GPT fragment can live in a smaller space than naively expected. Finally, it would be valuable to develop quantitative measures of the confidence in nonclassicality assessments obtained from GPT shadows, which would complement the qualitative consistency checks provided here by our theoretical model and its classical counterpart.

This work establishes integrated photonics as a platform for operational nonclassicality certification under minimal theoretical assumptions. More broadly, it provides a route for connecting experimentally accessible quantum technologies with certification methods whose conclusions do not depend on the prior validity of a specific physical theory.

\section*{Acknowledgments}

We thank Elie Wolfe for bringing the experimental and theoretical groups together at the outset of this project, and Albert Aloy for the support with the theory-agnostic tomography program. J.H.S. thanks Shane Mansfield for helpful discussions on integrated photonic circuits. V.P.R. acknowledges partial support by the Digital Horizon Europe project FoQaCiA, Foundations of quantum computational advantage, GA No.~101070558, funded by the European Union, NSERC (Canada), and UKRI (UK).
J.H.S. was funded by the European Commission by the QuantERA project ResourceQ under the grant agreement UMO2023/05/Y/ST2/00143. 
D.S. was supported by Perimeter Institute for Theoretical Physics. Research at Perimeter Institute is supported in part by the Government of Canada through the Department of Innovation, Science and Economic Development and by the Province of Ontario through the Ministry of Colleges and Universities. 
This work is partially carried out under the IRA Programme, project no.~FENG.02.01-IP.05-0006/23, financed by the FENG program 2021-2027, Priority FENG.02, Measure FENG.02.01., with the support of the FNP.
We acknowledge the use of a computational server financed by the Foundation for Polish Science (IRAP project, ICTQT, contract no. 2018/MAB/5/AS-1, co-financed by EU within Smart Growth Operational Programme). This work was supported by the Australian Research Council; E.P. is a recipient of an Australian Research Council Discovery Early Career Researcher Award (DE250100762). B.P., V.C., R.O., and F.S. acknowledge MUR PNRR project Spoke 4 and Spoke 7 (Grant No. PE0000023-NQSTI) and the ERC Advanced Grant QU-BOSS (QUantum advantage via non-linear BOSon Sampling, Grant No. 884676). Some figures were prepared using Mathcha.

\bibliographystyle{quantum}
\bibliography{biblio}

\appendix

\section{Rank analysis details}\label{ap:rank}

As stated in the main text, the standard theory-agnostic tomography analysis consists of, given a data table $D$ and a rank $k$, finding a factorization $S^T\cdot E$ that best describes $D$, such that the columns of $S$ and $E$ are valid states/effects for a GPT in $\mathbb{R}^k$. The decision of which model best fits $D$ is made by optimizing the parameter
\begin{equation}
    \chi^2(k,D) :=\min_{S,E}\sum_{ij}\frac{(S_i^T\cdot E_j-D_{ij})^2}{(\Delta D_{ij})^2},
\end{equation}
where $S_i$ and $E_j$ are the $i$-th column of $S$ and $j$-th column of $E$, respectively, and $(\Delta D_{ij})^2$ is the variance associated with the frequency provided by entry $D_{ij}$.

Given a collection of data tables $\{D^a\}_{a=1}^A$ obtained from the same experiment, one might leverage their statistical fluctuations to assess the rank. To do so, one obtains the parameter $\chi^2(k, D^a)$ as a function of the rank for each single data table. The average over the $A$ tables is referred to as $\frac{1}{A}\sum_a\chi^2(k,D^a)=:\chi^2_{\mathrm{train}}$. Then, one estimates the parameter $\chi^2$ that describes how well the optimal model obtained by theory-agnostic tomography for a given data table $D^a$ can describe the statistics of another data table $D^{b\neq a}$. That is, for $S^a$ and $E^a$ obtained as the optimal ones for $D^a$ and rank $k$, one computes
\begin{equation}
    \chi_{b\neq a}^2(k,D^a, D^b):=\sum_{ij}\frac{((S^a_i)^T\cdot E^a_j-D^b_{ij})^2}{(\Delta D^b_{ij})^2}\,.
\end{equation}
This is calculated for all valid pairings of data tables, and we refer to the average parameter as $\frac{1}{A(A-1)}\sum_{a,b}\chi_{b\neq a}^2=:\chi^2_{\mathrm{test}}$. The optimal rank $k$ is the one that simultaneously minimizes $\chi_{\mathrm{train}}^2$ and $\chi_{\mathrm{test}}^2$. The reason is that $\chi_{\mathrm{train}}^2$ will normally decrease monotonically with the rank, so the larger the rank, the smaller this parameter, and for most experiments, one would not find an upper bound for the rank by analyzing only $\chi_{\mathrm{train}}^2$. However, a rank too large will allow $\chi_{\mathrm{test}}^2$ to increase, since the individual theory-agnostic output for a particular data table $D^a$ will now be describing even the particular statistical fluctuations of this data, increasing $\chi_{b\neq a}^2$ when trying to describe the data of another table $D^{b\neq a}$. In other words, the selected rank corresponds to the best trade-off between accurately reproducing the observed data and maintaining the ability to generalize across independent realizations of the same experiment.

\section{Numerical implementation}\label{app:codes}

The numerical data for this work are available at Ref.~\cite{github}. This repository contains a Jupyter notebook with all the tools necessary to reproduce the results in this paper. We use the same algorithm as the one available in Ref.~\cite{aloy2024github}, with minor tweaks to accommodate the particular needs of this project. The specifics of how this program works can be found in Ref.~\cite{aloy2024theory}, and in this section (and notebook) we only summarise how the functions implement the routines from Ref.~\cite{aloy2024github}. The repository also calls functions from Ref.~\cite{cavalcanti2023github}. Please see Ref.~\cite{selby2024linear} for an explanation of how assessing simplex-embeddability of GPTs can be framed as a linear program.

The notebook is divided into three sections. The first defines and briefly explains the functions in Ref.~\cite{cavalcanti2023github} for assessing simplex-embeddability. Given a set of states and effects, the pipeline identifies the accessible GPT fragment~\cite{selby23fragments}, computes the facets of the state and effect positive cones via exact arithmetic (using the library \texttt{pycddlib}), and solves a linear program to obtain the robustness of contextuality $r\in[0,1]$. 

The second section implements the pipeline for rank analysis and theory-agnostic tomography, following Ref.~\cite{aloy2024github}. Given a frequency data table, it performs a seesaw optimisation over states and effects to minimize $\chi^2$ and computes the train and test errors across a range of ranks to identify the optimal factorization rank $k^*$ and construct the corresponding GPT fragment.

The third section provides the main functions employed in this paper. First, we define the function $\texttt{PTritterTheo(V\_P,V\_m,arm,outcome)}$ that reproduces the theoretical probabilities from the model introduced in Section~\ref{sec:setup} and detailed in Appendix~\ref{app:models}. From this function and from the experimental data (also available in the repository), we provide functions to generate data tables that mimic these statistics under Poissonian noise for a given total number of coincidences \texttt{N}. This section also provides the commands employing the pipelines from the previous sections to perform rank analysis, theory-agnostic tomography, and the simplex-embeddability check.

The repository also contains files with all the data tables employed in this manuscript and their respective rank analyses, as well as the file containing the experimental data. The linear program in section 1 is solved using the \texttt{MOSEK} solver via \texttt{cvxpy}. An academic licence for \texttt{MOSEK} is freely available. The quadratic programs in section 2 are solved using \texttt{cvxopt}. All other dependencies (\texttt{numpy}, \texttt{scipy}, \texttt{sympy}, \texttt{matplotlib}) are standard and freely available via \texttt{pip}. Our computations were run on an HPE ProLiant server with an AMD EPYC 7713 64-core processor at 2.4 GHz and 1 TB of RAM, running RHEL 8.1.

\section{Theoretical model parameters}\label{app:models}

\begin{figure}[b]
    \centering  \adjustbox{width=\linewidth,valign=t}{\input{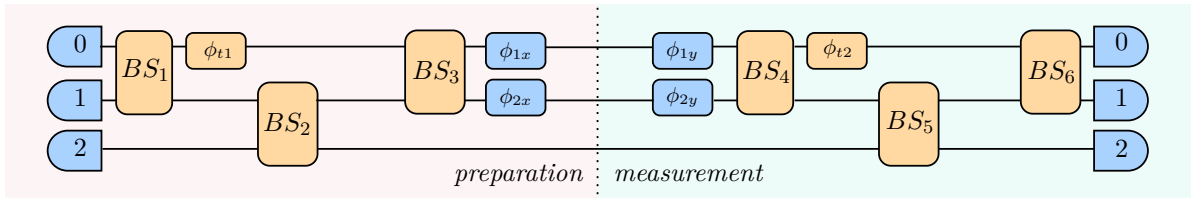}}
    \caption{Detailed schematic representation of the theoretical model describing the photonic circuit. Blue components are tunable during the experiment, while yellow components are fixed throughout all runs.}
    \label{fig:circuit-detailed}
\end{figure}

As introduced in Section~\ref{sec:setup}, we employ a theoretical model to describe the experimental data obtained from the photonic processor. The detailed schematic representation of this model is provided in Figure~\ref{fig:circuit-detailed}. The values of all parameters mentioned below are provided in Table~\ref{tab:params}. The input states are described by elements of a three-mode Hilbert space,
\begin{equation}
    \ket{0}=\begin{pmatrix}
        1\\0\\0
    \end{pmatrix},\quad\ket{1}=\begin{pmatrix}
        0\\1\\0
    \end{pmatrix},\quad\ket{2}=\begin{pmatrix}
        0\\0\\1
    \end{pmatrix}.
\end{equation}

Beam-splitters $BS_1$, $BS_3$, $BS_4$ and $BS_6$ have the form
\begin{equation}\label{eq:BS1}
    BS_j=\begin{pmatrix}
        r_j & it_j & 0\\
        it_j & r_j & 0\\
        0 & 0 & 1
    \end{pmatrix},\quad j\in\{1,3,4,6\},
\end{equation}
while beam-splitters $BS_2$ and $BS_5$ are given by
\begin{equation}\label{eq:BS2}
    BS_j=\begin{pmatrix}
        1 & 0 & 0\\
        0 & r_j & it_j\\
        0 & it_j & r_j
    \end{pmatrix},\quad j\in\{2,5\},
\end{equation}
with $t_j=\sqrt{1-r^2_j}$ and $r_j$ provided in Table~\ref{tab:params} for all values of $j$. The fixed phase-shifters $\hat{\phi}_{t1}$ and $\hat{\phi}_{t2}$ are described by
\begin{equation}\label{eq:PS1}
    \hat{\phi}_{tj}=\begin{pmatrix}
        e^{i\phi_{tj}} & 0 & 0\\
        0 & 1 & 0\\
        0 & 0 & 1
    \end{pmatrix},\quad j\in\{1,2\}
\end{equation}
while $\hat{\phi}_{02j}$, with $j=1,2,3,4$, are described by
\begin{equation}\label{eq:PS2}
    \hat{\phi}_{02j}=\begin{pmatrix}
        1 & 0 & 0\\
        0 & e^{i\phi_{02j}} & 0\\
        0 & 0 & 1
    \end{pmatrix}, \quad j\in\{1,2,3,4\}
\end{equation}

As detailed in Sec.~\ref{sec:setup}, the phase shifts modulated by the potentials $\Delta V_x$ and $\Delta V_y$ can in principle be nonlinear. The potentials themselves require corrections that take into account internal resistances of the photonic setup, such that the quantities that effectively tune $\phi_1$ and $\phi_2$ are given by the dissipated powers
\begin{equation}\label{eq:Veff}
    P(\Delta V_x):=\frac{a_1 \Delta V_x^2}{1+b_1\Delta V_x^2},\quad P(\Delta V_y):=\frac{a_2\Delta V_y^2}{1+b_2\Delta V_y^2}.
\end{equation}
 The phases $\phi_1$ and $\phi_2$ are given as functions of $P(\Delta V_x)$ and $P(\Delta V_y)$ as

\begin{equation}\label{eq:phi1}
\begin{split}
    \phi_1(\Delta V_x,\Delta V_y)&=\alpha_{11}P(\Delta V_x)+\alpha_{112}P(\Delta V_x)^2+ \alpha_{21}P(\Delta V_y)+\alpha_{212}P(\Delta V_y)^2+\phi_{01},
    \end{split}
\end{equation}
\begin{equation}\label{eq:phi2}
\begin{split}
    \phi_2(\Delta V_x,\Delta V_y)&=\alpha_{12}P(\Delta V_x)+\alpha_{122}P(\Delta V_x)^2+\alpha_{22}P(\Delta V_y)+\alpha_{222}P(\Delta V_y)^2+\phi_{02},
    \end{split}
\end{equation}
with parameter values given in Table~\ref{tab:params}. This model reproduces the experimental data (see Fig.~\ref{fig:tritter}), and allows us to define the phases
\begin{equation}
    \phi_1^x(\Delta V_x):=\alpha_{11}P(\Delta V_x)+\alpha_{112}P(\Delta V_x)^2;\quad\phi_1^y(\Delta V_y):=\alpha_{21}P(\Delta V_y)+\alpha_{212}P(\Delta V_y)^2;
\end{equation}
\begin{equation}
    \phi_2^x(\Delta V_x):=\alpha_{12}P(\Delta V_x)+\alpha_{122}P(\Delta V_x)^2;\quad\phi_2^y(\Delta V_y):=\alpha_{22}P(\Delta V_y)+\alpha_{222}P(\Delta V_y)^2.
\end{equation}

The probabilities are thus given by the squared transition amplitude between the effect
\begin{equation}
    \bra{k,\Delta V_y}:=\bra{k}BS_6\cdot\hat{\phi}_{024}\cdot BS_5\cdot \hat{\phi}_{023}\cdot\hat{\phi}_{t2}\cdot BS_4\cdot \hat{\phi}_2^y(\Delta V_y)\cdot\hat{\phi}_1^y(\Delta V_y)
\end{equation}
and the state
\begin{equation}
    \ket{0,\Delta V_x}:=\hat{\phi}^x_2(\Delta V_x)\cdot\hat{\phi}^x_1(\Delta V_x)\cdot BS_3\cdot\hat{\phi}_{022}\cdot BS_2\cdot\hat{\phi}_{021}\cdot\hat{\phi}_{t1}\cdot BS_1\ket{0},
\end{equation}
such that
\begin{equation}
    p(k|\Delta V_y,\Delta V_x,0)=|\!\braket{k,\Delta V_y|0,\Delta V_x}\!|^2.
\end{equation}

We can moreover construct the set of states $\Omega:=\{\ket{0,\Delta V_x}\!\!\bra{0,\Delta V_x}\}_{\Delta V_x}$ and effects $\mathcal{E}:=\{\ket{k,\Delta V_y}\!\!\bra{k,\Delta V_y}\}_{\Delta V_y,\,k=0,1,2}$ from this model. As discussed in Sec.~\ref{sec:results}, while the dimension of $\mathsf{LinSpan}(\mathcal{E})$ is equal to $9$ (i.e., the full dimension of this GPT), we have a smaller dimension for $\mathsf{LinSpan}(\Omega)$ equal to 7. This coincides with the optimal rank for theory-agnostic tomography for both this model and the experimental data. The reason why we do not pass $(\Omega,\mathcal{E})$ straight to the linear program~\cite{cavalcanti2023github} is twofold: on the one hand, the jump from dimension 7 to 9 is computationally demanding for polytope representation transformation, which is a crucial step in the program. In fact, it has been argued that for numerical techniques similar to our simplex-embeddability assessment, the complexity grows exponentially with the dimension of the polytope~\cite{sahandeh26complexity}. On the other hand, performing theory-agnostic tomography on the theoretical data allows for a more direct comparison with the results obtained from the experimental implementation.

\begin{table}
    \centering
    \begin{tabular}{ccc}\hline
        Label & Value & Equation \\\hline
        $r_1$ & $\sqrt{\frac12-0.001311}$ & \ref{eq:BS1}\\
        $r_2$ & $\sqrt{\frac13-0.066085}$ & \ref{eq:BS2}\\
        $r_3$ & $\sqrt{\frac12-0.000988}$ & \ref{eq:BS1}\\
        $r_4$ & $\sqrt{\frac12+0.022457}$ & \ref{eq:BS1}\\
        $r_5$ & $\sqrt{\frac13-0.082587} $& \ref{eq:BS2}\\
        $r_6$ & $\sqrt{\frac12+0.033388}$  & \ref{eq:BS1}\\\hline
        $\phi_{t1}$ & $1.270043$ & \ref{eq:PS1}\\
        $\phi_{t2}$ & $0.674953$ & \ref{eq:PS1}\\
        $\phi_{021},\phi_{022},\phi_{023},\phi_{024}$ & 0 & \ref{eq:PS2}\\\hline
        $a_1$ & $83.011984$ & \ref{eq:Veff}\\
        $b_1$ & $94.140345$ & \ref{eq:Veff}\\
        $a_2$ & $105.366774$ & \ref{eq:Veff}\\
        $b_2$ & $80.800415$ & \ref{eq:Veff}\\\hline
        $\alpha_{11}$ & $30.084171$ & \ref{eq:phi1}\\
        $\alpha_{112}$ & $73.845614$ & \ref{eq:phi1}\\
        $\alpha_{21}$ & $29.908311$ & \ref{eq:phi1}\\
        $\alpha_{212}$ & $43.721934$ & \ref{eq:phi1}\\
        $\phi_{01}$ & $0.273968$ & \ref{eq:phi1}\\
        $\alpha_{12}$ & $10.229360$ & \ref{eq:phi2}\\
        $\alpha_{122}$ & $21.886606$ & \ref{eq:phi2}\\
        $\alpha_{22}$ & $8.545192$ & \ref{eq:phi2}\\
        $\alpha_{222}$ & $10.70093$ & \ref{eq:phi2}\\
        $\phi_{02}$ & $0.029033$ & \ref{eq:phi2}
        \ref{eq:phi1},\ref{eq:phi2}\\\hline
    \end{tabular}
    \caption{Parameters for the theoretical model simulating the experimental statistics, obtained by fitting to the data.}
    \label{tab:params}
\end{table}

\section{Consistency check}\label{app:sanity}

\begin{figure}[htb!]
    \centering
    \adjustbox{width=0.75\linewidth,valign=t}{\input{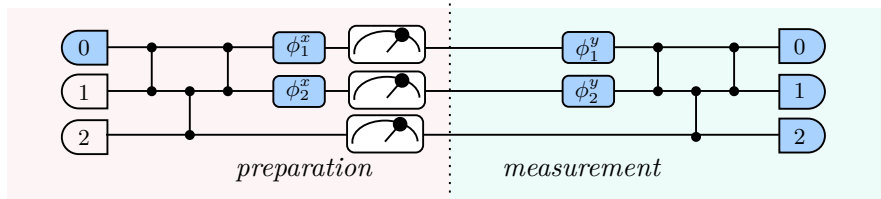}}
    \caption{A classical counterpart to the photonic experiment, where the preparations undergo full dephasing before the measurement procedure. The parameters are the same as the ones employed in the theoretical model (see Appendix~\ref{app:models}).}
    \label{fig:dephased-circuit}
\end{figure}

In order to further gather evidence that the nonclassicality certified by the analysis employed in this work is genuine, we investigate whether the theory-agnostic tomography analysis preserves classicality. As previously established, it is an open question whether the assessment of nonclassicality over a GPT constructed via theory-agnostic tomography is trustworthy~\cite{schmid2025shadows}. We therefore consider the data obtained from a classical experiment, in which a fully dephasing channel is introduced before the preparation is sent to the measurement apparatus. The remainder of the model is kept exactly the same as the one employed in the main text, and described in Appendix~\ref{app:models}. A scheme of such a scenario is provided in Fig.~\ref{fig:dephased-circuit}.

\begin{figure}
    \centering
    \begin{tabular}{cc}
    A) \adjustbox{valign=t}{\includegraphics[width=0.45\linewidth]{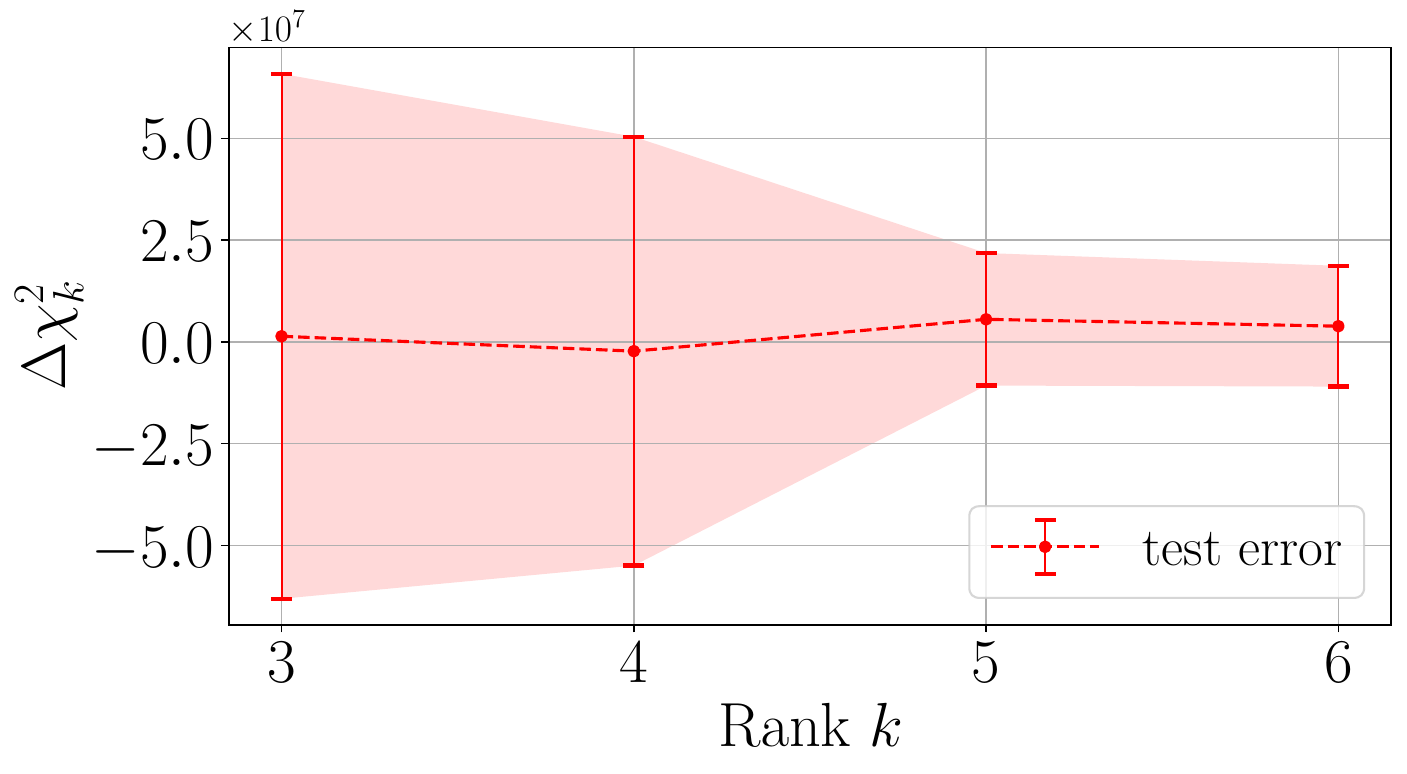}} &  F) \adjustbox{valign=t}{\includegraphics[width=0.45\linewidth]{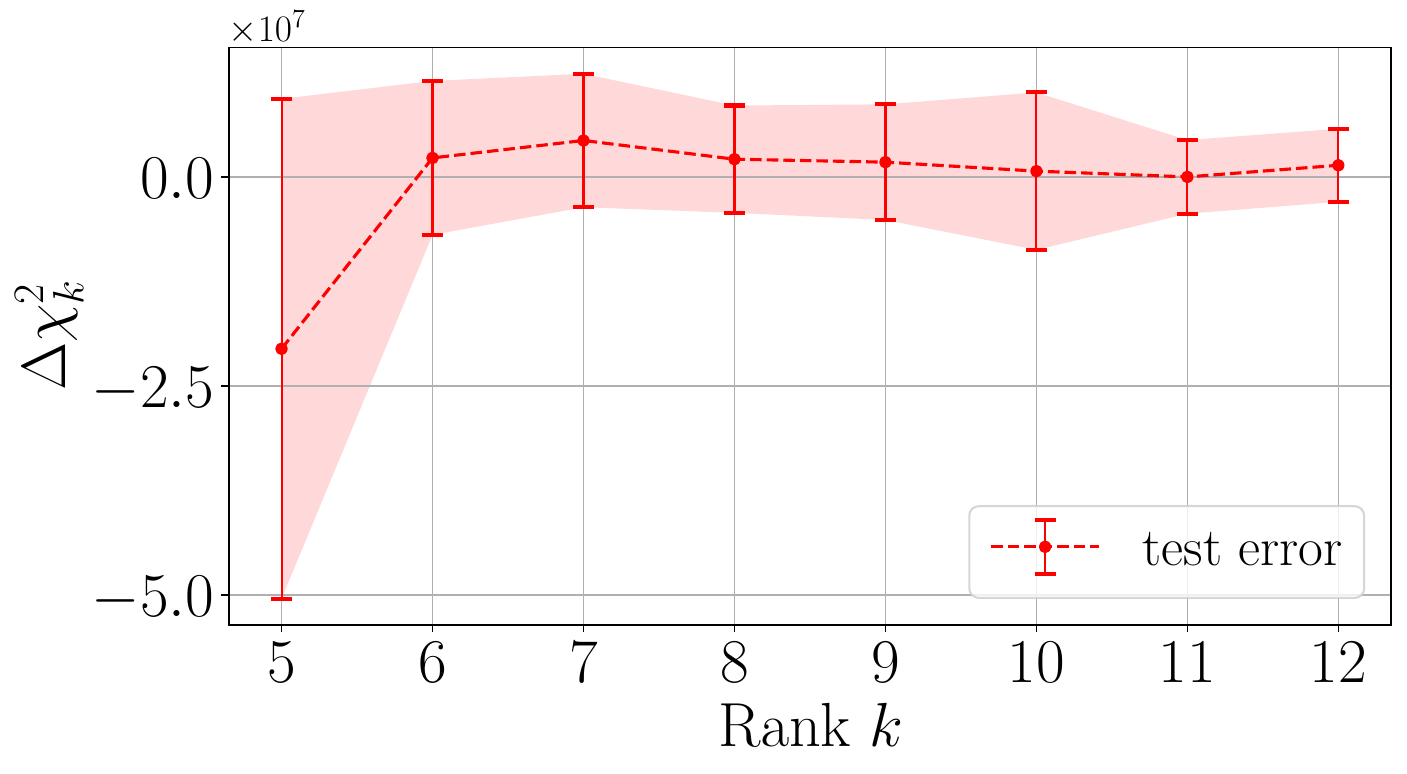}}\\
    B)\adjustbox{valign=t}{\includegraphics[width=0.45\linewidth]{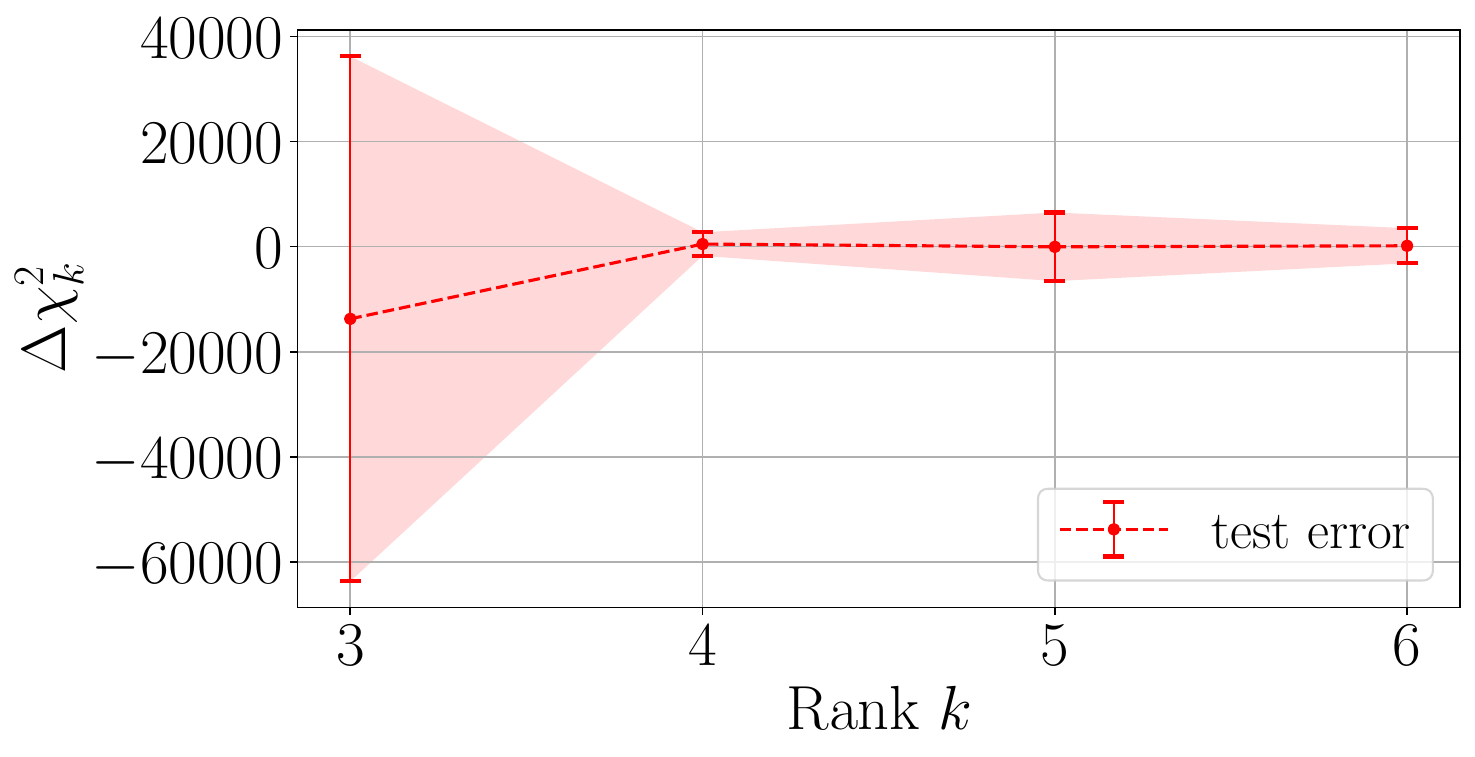}} & G)\adjustbox{valign=t}{\includegraphics[width=0.45\linewidth]{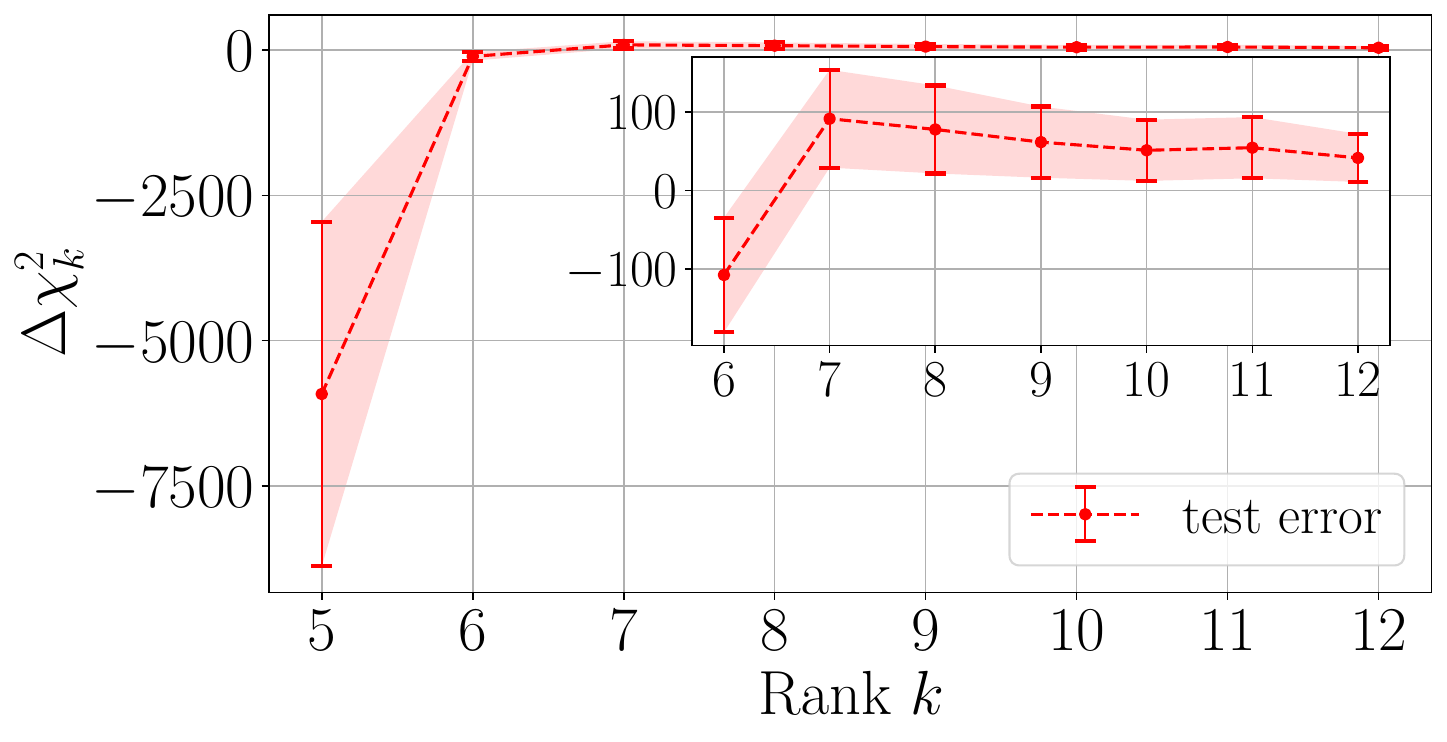}} \\
    C)\adjustbox{valign=t}{\includegraphics[width=0.45\linewidth]{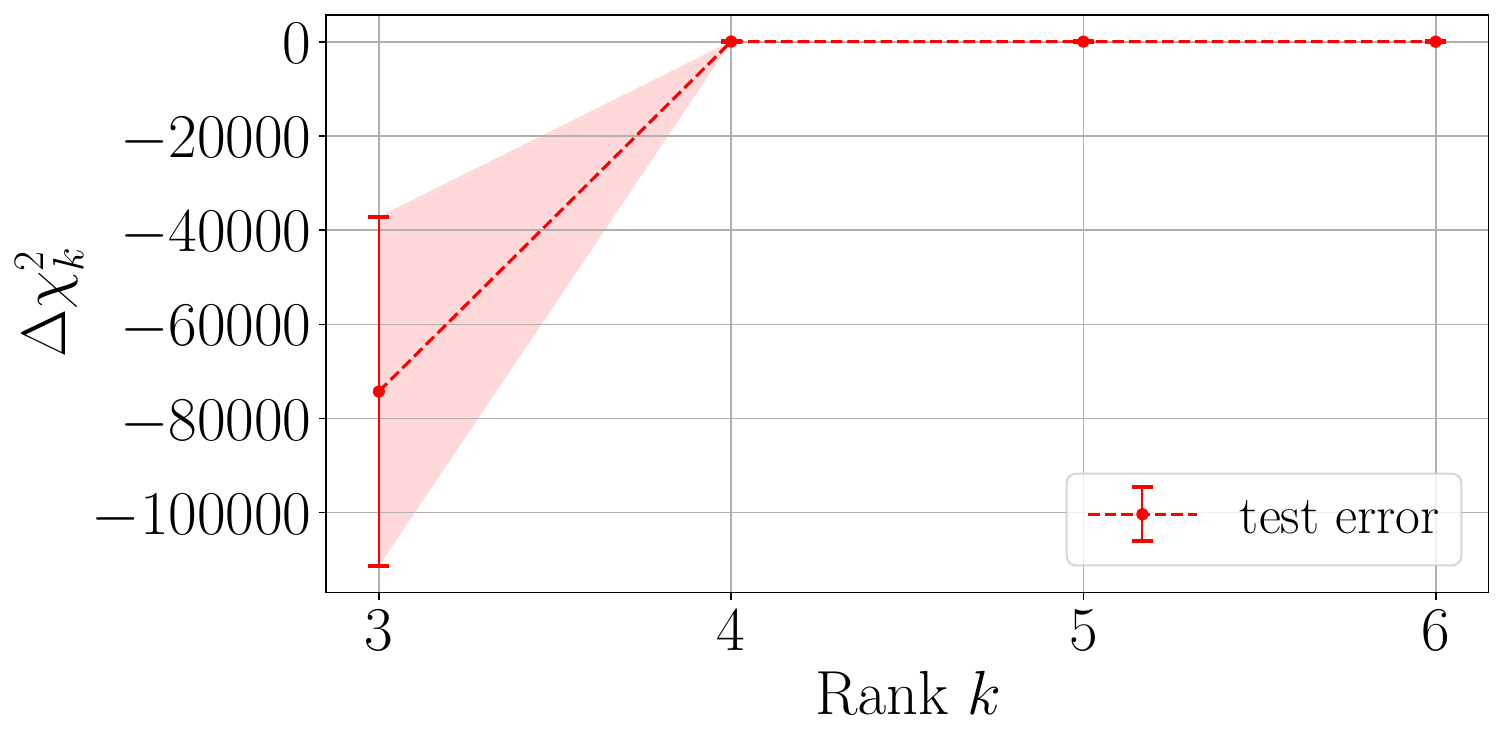}} &  H) \adjustbox{valign=t}{\includegraphics[width=0.45\linewidth]{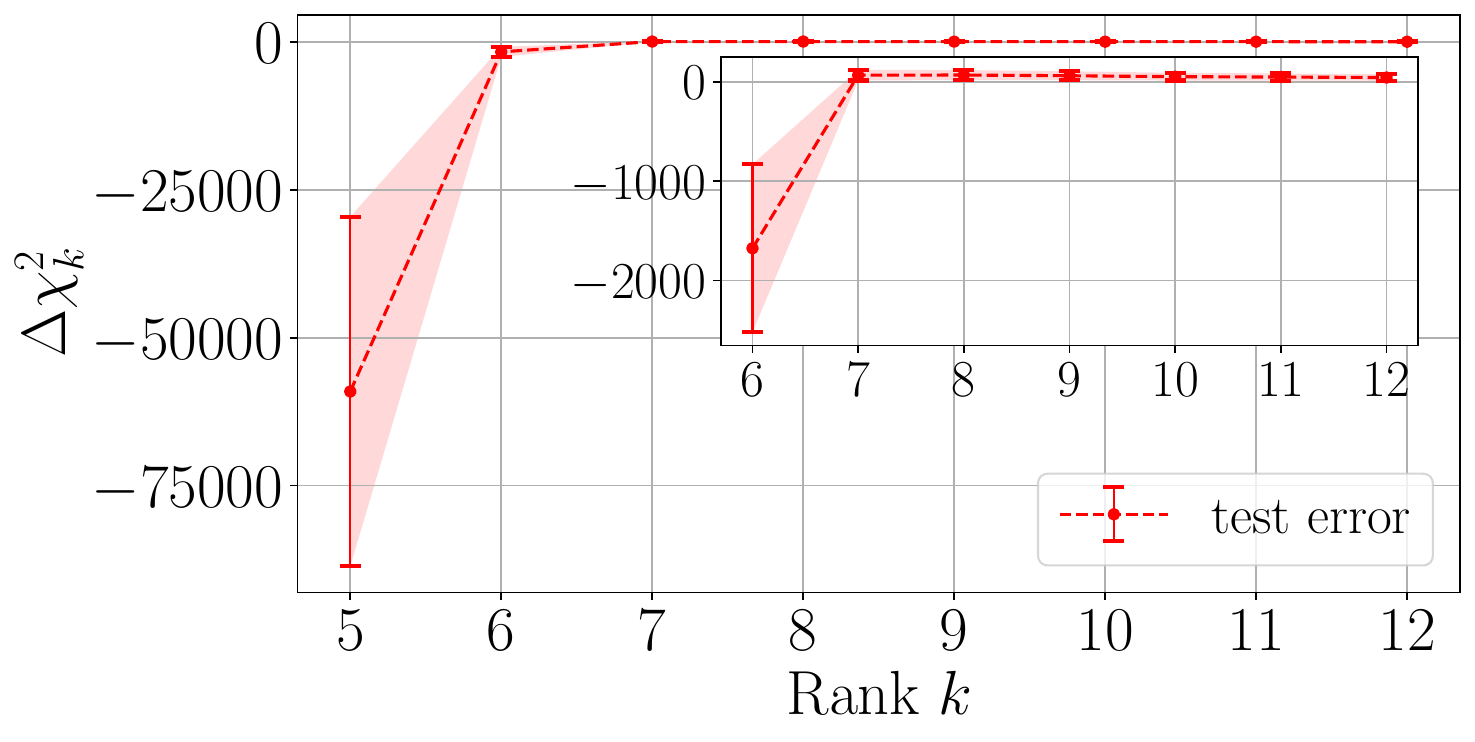}}\\
    D)\adjustbox{valign=t}{\includegraphics[width=0.45\linewidth]{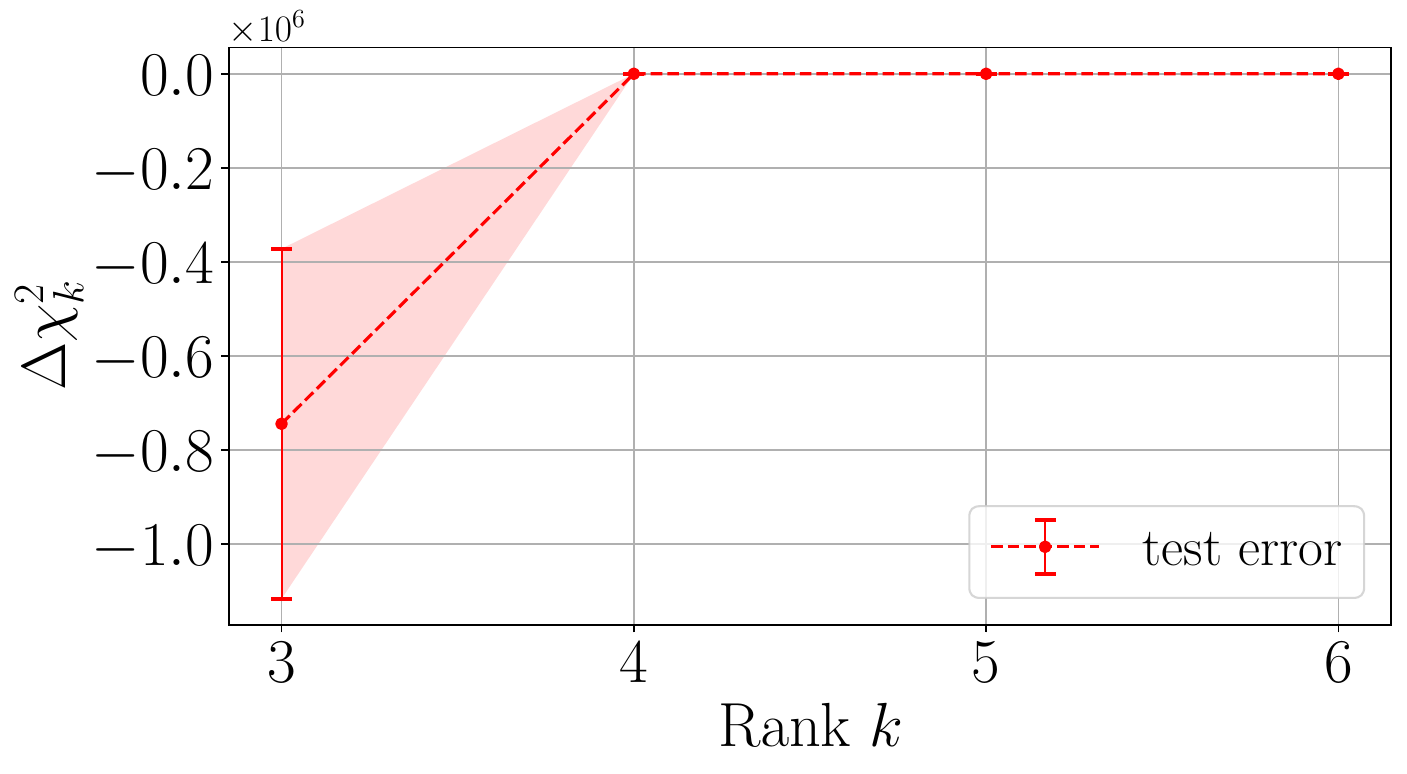}} &  I) \adjustbox{valign=t}{\includegraphics[width=0.45\linewidth]{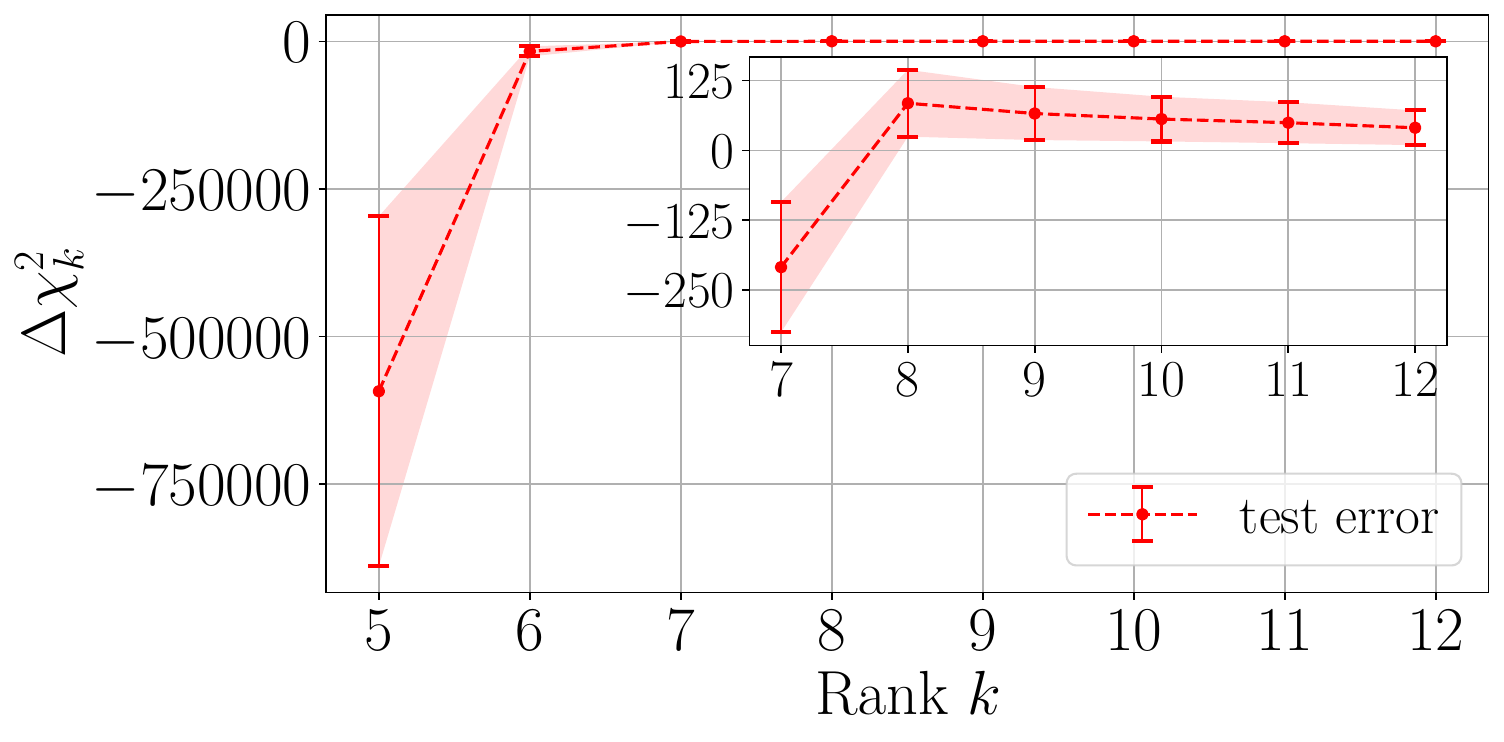}}\\
    E)\adjustbox{valign=t}{\includegraphics[width=0.45\linewidth]{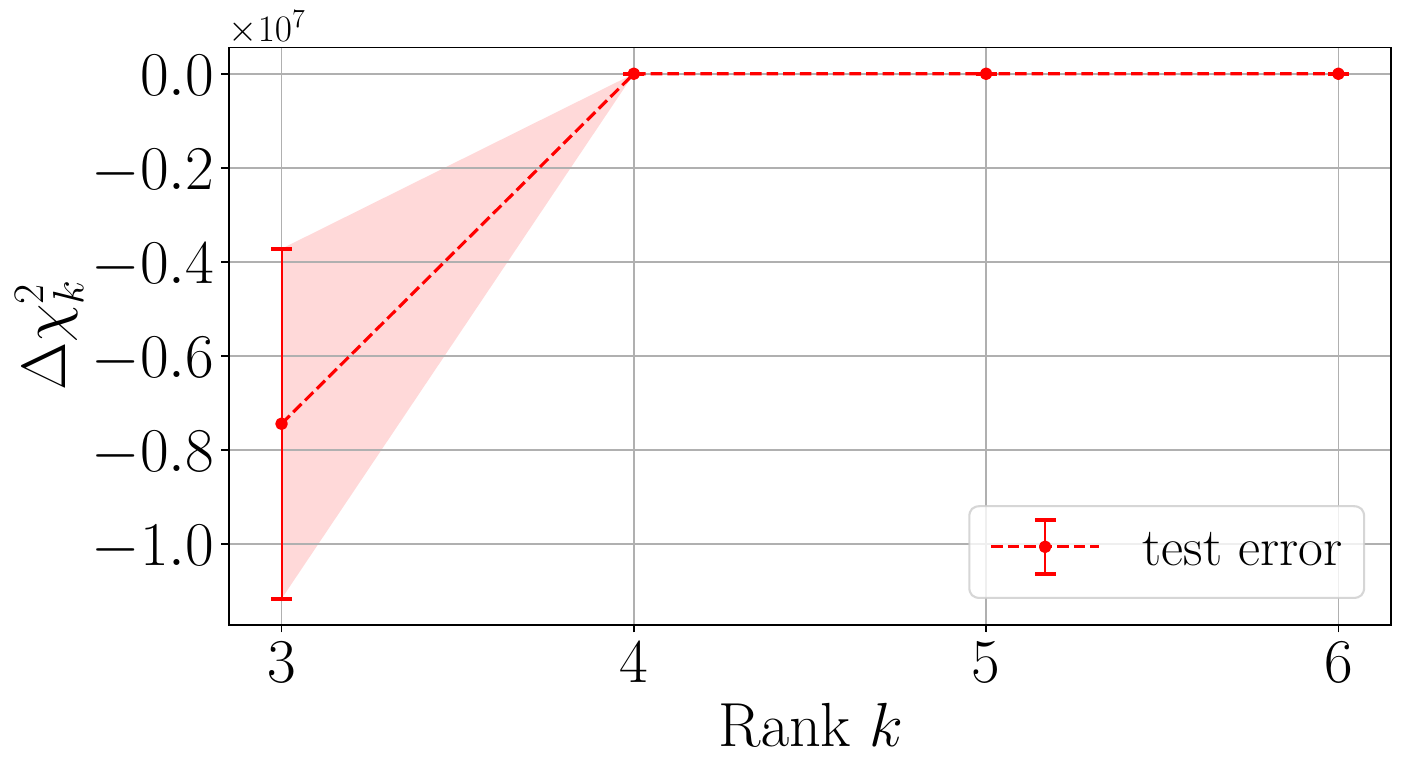}} &  J) \adjustbox{valign=t}{\includegraphics[width=0.45\linewidth]{deltachisq_theo4.pdf}}
    \end{tabular}
    \caption{Change in $\chi^2_{\mathrm{test}}$ for the data obtained from Poissonian sampling of the statistics of (A--E) the classical circuit in Fig.~\ref{fig:dephased-circuit} and (F--J) the theoretical model introduced in Appendix~\ref{app:models}, assuming the number of coincidences to be (A,F) $N=10^2$, (B,G) $N=10^3$, (C,H) $N=10^4$, (D,I) $N=10^5$, (E,J) $N=10^6$. For the dephased model one has optimal rank $k=2$ for $N=10^2$--$10^3$ and $k=3$ for all other values, whereas for the full theoretical model one has $k=4$ for $N=10^2$, $k=6$ for $N=10^3$--$10^4$, and $k=7$ for $N=10^5$--$10^6$.}
    \label{fig:dephased-rank}
\end{figure}

We can generate the density matrices and POVM elements directly from the theoretical model and run them through the linear program~\cite{cavalcanti2023github}, obtaining the expected robustness of contextuality $r=0$. This certifies that the theory-dependent ideal data is classical. We then move on to performing a theory-agnostic analysis of the data tables that these density matrices and POVM elements generate. For this, we perform a rank analysis in a similar way to the one performed for the theoretical model in Sec.~\ref{se:rank}---that is, we take not only the ideal data table generated by the analytical states and effects, but also modifications of it obtained by applying Poissonian noise. For this analysis, we assume the total number of coincidences to be $N=10^6$, and we sample 9 additional data tables from these statistics via multinomial sampling. The rank $k=3$ is obtained as the optimal one, as expected. The analysis can be found in Fig.~\ref{fig:dephased-rank}. When we finally feed the theory-agnostic output to the linear program, not only the original data but also the Poissonian fluctuations are all assessed to be classical---that is, they feature a robustness $r=0$ up to numerical accuracy. This further builds the case that the nonclassicality observed in the experiment is genuine and does not stem from classical noise.

\begin{figure}
    \centering
    \includegraphics[width=0.55\linewidth]{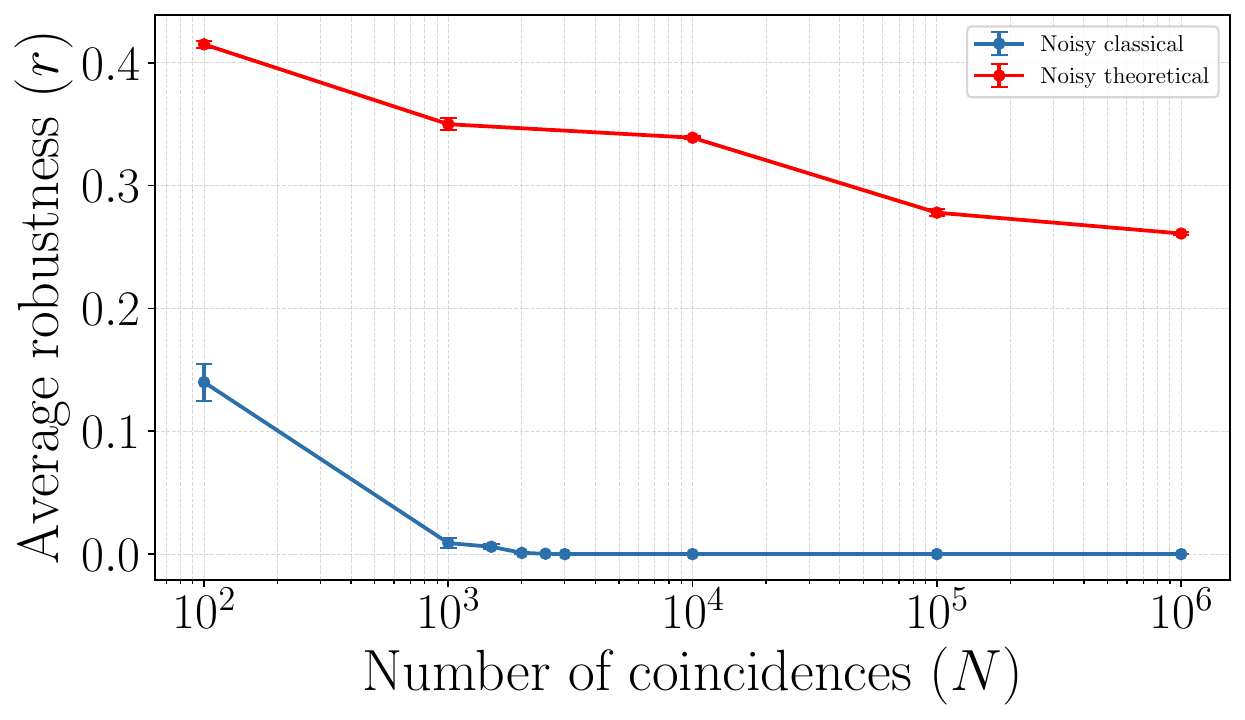}
    \caption{Average robustness of contextuality for the data obtained from Poissonian sampling of the statistics of the classical circuit in Fig.~\ref{fig:dephased-circuit} (blue) and of the theoretical model describing the experiment (red), assuming $N$ coincidences. Notice that the analysis leads to a correct assessment of classicality as long as the number of coincidences lies above $N=2\cdot 10^3$.}
    \label{fig:dephased-N}
\end{figure}

We can push the analysis further to investigate how low the number of coincidences $N$ can be while still obtaining a sound assessment. We therefore consider different values of $N$ and implement the same analysis, generating the fluctuation data tables and checking for the optimal rank. This analysis is compiled in Fig.~\ref{fig:dephased-rank}. We can see that for all values of $N$, the optimal rank is always $k=3$. We then provide the rank analysis output for all the data tables to the linear program, and compile the average robustness values in Fig.~\ref{fig:dephased-N}. One can see that, when the number of counts is around $N=2\cdot10^3$ or below, we start getting nonzero robustness for some of the noisy outputs. For the range considered in this analysis ($N\approx3.5\cdot10^3$ for the experimental data), however, the assessment holds for all noisy data tables.

\end{document}